\documentclass[12pt, a4paper]{article}
\usepackage[T1]{fontenc}
\usepackage[polish]{babel}
\addto\captionspolish{}
\usepackage{algorithm}
\usepackage{algpseudocode}
\usepackage{amsmath}
\numberwithin{equation}{section}
\usepackage{amssymb}
\RequirePackage{amsthm}
\usepackage{array}
\usepackage{authblk}
\usepackage{blkarray}
\usepackage{booktabs}
\usepackage{caption}
\usepackage{chngcntr}
\usepackage{comment}
\usepackage{dsfont}
\usepackage{enumerate}
\usepackage{enumitem}
\usepackage{eurosym}
\usepackage{geometry}
\usepackage{graphicx}
\usepackage[colorlinks=true, allcolors=blue]{hyperref}
\usepackage{lscape}
\usepackage{mathrsfs}
\usepackage{mathtools}
\usepackage{natbib}
\usepackage{nicematrix} 
\usepackage{tikz}
\usepackage{upquote}

\addto\captionspolish{

}

\theoremstyle{plain}

\counterwithin{algorithm}{section}

\newtheorem{As}{Assumption}[section]
\newtheorem{Cor}{Corollary}[section]

\newtheorem{Prop}{Proposition}[section]
\newtheorem{Rem}{Remark}[section]
\newtheorem{Theo}{Theorem}[section]

\theoremstyle{remark}
\newtheorem{Def}{Definition}[section]
\newtheorem{Ex}{Example}[section]

\numberwithin{Alg}{section}
\numberwithin{As}{section}
\numberwithin{Cor}{section}
\numberwithin{Def}{section}
\numberwithin{Ex}{section}
\numberwithin{Lem}{section}
\numberwithin{Prop}{section}
\numberwithin{Rem}{section}
\numberwithin{Theo}{section}

\DeclareMathOperator{\Exp}{\mathds E}
\DeclareMathOperator{\Prob}{\mathds P}

\begin{document}

\title{Functional Periodic ARMA Processes}

\author[1]{Sebastian K\"uhnert}
\author[2]{Juhyun Park}

\affil[1]{Department of Mathematics, Ruhr University Bochum, 44780 Bochum, Germany, \textsuperscript{*}\href{mailto:sebastian.kuehnert@ruhr-uni-bochum.de}{sebastian.kuehnert@ruhr-uni-bochum.de}}
\affil[2]{Laboratoire de Mathématiques et Modélisation d'Evry, Université Paris-Saclay, ENSIIE, UEVE, UMR 8071 CNRS, France, \textsuperscript{\dag}\href{mailto:juhyun.park@ensiie.fr}{juhyun.park@ensiie.fr}}

\date{December 16, 2025}

\maketitle

\begin{abstract} Periodicity is a common feature of time series. For finite-dimensional data, periodic autoregressive moving average (ARMA) models have been extensively studied. In functional time series analysis, AR  models have been extended to incorporate periodicity, but existing approaches remain incomplete and do not cover the ARMA setting. This paper develops a rigorous theoretical framework for functional periodic ARMA (fPARMA) processes in general separable Hilbert spaces. The proposed model class accommodates periodically varying dependence structures. We derive sufficient conditions for periodic stationarity, the existence of finite moments, and weak dependence. Moreover, we study Yule-Walker-type estimators for the fPAR operators and, in a specific setting, estimators for the fPARMA operators, and establish convergence rates under Sobolev-type regularity assumptions.
\end{abstract}

\noindent{\small \textit{MSC 2020 subject classifications:} 62F12, 62M10, 62R10}

\noindent{\small \textit{Keywords:} AR; ARMA; functional data; functional time series; PARMA; periodicity}

\section{Introduction} 

Functional data analysis has grown substantially over recent decades, driven by the increasing availability of complex, high-dimensional data observed over continuous domains such as time, space, or frequency. This development has stimulated the creation of theoretical and methodological tools specifically tailored to the functional setting. Several foundational contributions have laid the groundwork for this field; see, e.g., \citet{FerratyVieu2006,HorvathKokoszka2012,HsingEubank2015,RamsaySilverman2005}. In many real-world applications, functional observations are recorded sequentially over time, giving rise to the field of functional time series analysis, where we refer to  \cite{Bosq2000, kokoszka:2017:FDA-book} for comprehensive overviews. The standard framework assumes that such observations take values in a separable Hilbert space, most commonly $L^2[0,1]$, the space of square-integrable functions on the unit interval. While trivial cases include deterministic sequences or independent (possibly i.i.d.) observations, the main focus of the literature lies on stationary yet serially dependent functional time series \citep[for procedures on testing stationarity, see][]{HorvathKokoszkaRice2014,AueVanDelft2020,vanDelftCharaciejusDette2021}. Prominent applications arise in economics \citep{Portelaetal2017,Riceetal2023}, electricity demand analysis \citep{Chaouch2013}, and demography \citep{HyndmanShang2009}.

In close analogy to classical time series analysis, linear models play a central role in the functional setting. This includes functional autoregressive (fAR), functional moving average (fMA), and functional autoregressive moving average (fARMA) processes, which are widely used due to their tractability, interpretability, and rich theoretical properties. Under mild regularity conditions, such models admit strictly stationary solutions. Detailed treatments of fAR and fARMA processes and their statistical properties can be found in \citet{Bosq2000,Chenetal2021,KlepschKlueppelbergWei2017,Spangenberg2013}. Identifiability issues and consistent estimation procedures for fARMA models are studied in \citet{kuenzer:2024}, while further results on consistent estimation for fARMA models and functional invertible linear processes are provided in \citet{KuehnertRiceAue2025}.

Beyond stationarity, several forms and features of non-stationary functional time series have been investigated, including functional cointegration \citep{BeareSeoSeo2017,ChangMillerPark2017}, local stationarity \citep{vanDelftEichler2018}, structural breaks \citep[see][]{AueRiceSoenmez2017,BuecherDette2018}, presence of periodic (deterministic) components \citet{HoermannKokoszkaNisol2018}, and periodically correlated functional processes \citep{KidzińskiKokoszkaMohammadi2018}. While periodic behavior has received comparatively less attention in the functional literature, it is well studied in the scalar setting; see \citet{FransesPaap2004} and \citet{GardnerNapolitanoPaura2006} for comprehensive overviews, with applications in finance \citep{AndersenSuTodorovZhang2024} and the natural sciences \citep{Bloomfield_Hurd_Lund_1994,Lund_Hurd_Bloomfield_Smith_1995}.

Scalar periodic autoregressive (PAR) models were introduced in \cite{Pagano1978} and studied independently in \cite{Troutman1979}. A process $(X_k)$ is said to follow a PAR model of period $T\in\mathbb{N}$ if
\[
    X_k = \sum_{i=1}^{p(k)} \,\phi_{i,k} X_{k-i} + \varepsilon_k,
\]
for each $k$, where all coefficients $\phi_{i,k},$ the order $p(k),$ and the distribution of the innovations $\varepsilon_k$ depend on the season $s \in \{1,\dots,T\}$, with $s = (k \bmod T) + 1$.
An extension to periodic autoregressive moving average (PARMA) processes was firstly mentioned in \citet{ClevelandTiao1979}. In this case, $(X_k)$ follows a PARMA model of period $T$ if
\[
    X_k = \sum_{i=1}^{p(k)} \,\phi_{i,k} X_{k-i}
          + \sum_{j=1}^{q(k)} \,\psi_{j,k} \varepsilon_{k-j}
          + \varepsilon_k,
\]
for each $k$, where all coefficients $\phi_{i,k}$, $\psi_{j,k}$, the orders $p(k)$ and $q(k),$ and the distribution of the innovations $\varepsilon_k$ depend on the season. For estimation methods for PARMA models, \citet{adams1995,Francq_Roy_Saidi_2011,SarnagliaEtAl2016, Vecchia1985}, while procedures for detecting periodicity are discussed in \citet{KreissMaourisPaparoditis2025}. For extensions of PAR and PARMA models to the multivariate setting, see \citet{AknoucheHamdi2009,Bartolinietal1988,Rasmussenetal1996,Luetkepohl2005,DzikowskiJentsch2025} and the references therein.

A functional extension of the PAR model of order one, termed the functional periodic autoregressive (fPAR) model, was first proposed by \citet{SoltaniHashemi2011a} in the Hilbert space setting. In this framework, the process admits the representation
\begin{equation}\label{sk3}
    X_k = \phi_{1,k}(X_{k-1}) + \varepsilon_k,
\end{equation}
where $\phi_{1,k}$ are bounded linear operators and $\varepsilon_k$ are Hilbert space-valued innovations, both varying periodically. Higher-order extensions were subsequently proposed in \citet{SoltaniHashemi2011b}, albeit discussed only briefly.
Estimation procedures for the first-order fPAR model were studied in \citet{HashemiZamaniHaghbin2019}, and extensions to Banach space-valued processes were proposed by \citet{Parvardeh2017}. For a general discussion of periodically correlated functional time series, see \citet{KidzińskiKokoszkaMohammadi2018}, while consistent estimation of the period was addressed in \citet{CeroveckiCharaciejusHoermann2022}. We emphasize that a somewhat related but distinct strand of the literature concerns functional seasonal ARMA models \citep{Mestre2020} and functional seasonal AR processes as introduced in \citet{ZamaniHaghbinHashemiHyndman2022}. The latter take the form $X_k = \phi(X_{k-T}) + \varepsilon_k$, where $\phi$ and the distribution of the innovations $\varepsilon_k$ do not depend on the season, in contrast to fPAR models.

This article develops a rigorous theory of periodic models in the functional setting by extending the functional periodic autoregressive model in \eqref{sk3} to a functional periodic autoregressive moving average (fPARMA) framework in separable Hilbert spaces. The proposed model is driven by white noise innovations whose distributions vary periodically over time. Our framework allows both the autoregressive and moving average orders, as well as their associated operators, to depend on the season, thereby enabling flexible modeling of cross-seasonal dynamics. We formalize the notion of periodic stationarity (\emph{cyclostationarity}), its wide-sense counterpart, also referred to as \emph{periodic correlatedness}, and the concept of periodic white noise in the functional context. Exploiting the one-to-one correspondence between periodic and multivariate time series, where the period determines the vector dimension, we derive sufficient conditions for cyclostationarity via a multivariate reformulation. We further establish conditions for the existence of finite moments and weak dependence, which in turn ensure consistent estimation and, in specific settings, the validity of a central limit theorem. In addition, we provide explicit expressions for the lagged covariance operators of the proposed model. Further, in the autoregressive setting, we propose a practical estimation procedure together with an implementable algorithm. Assuming consistent estimation of the associated block operator matrix, the seasonal operators are recovered using Tikhonov-regularized pseudo-inverses. Under a Sobolev-type smoothness condition in the spirit of \cite{hall:meister:2007}, we establish convergence rates and illustrate the proposed methodology through extensive examples. Finally, we outline the estimation procedure and derive consistency rates for the fPARMA parameters under suitable regularity conditions.

The remainder of the article is organized as follows. Section~\ref{Section: Preliminaries} introduces the notation and the concepts of cyclostationarity, periodic correlatedness, and periodic white noise. Section~\ref{Section Model} defines the fPARMA model, presents its multivariate representation, and establishes its relation to functional ARMA models. Stationarity conditions and structural properties are derived in Section~\ref{sec:properties}. Estimation methods for fPAR operators are developed in Section~\ref{Section: Estimation fPAR}, while corresponding procedures for fPARMA operators are outlined in Section~\ref{Section: Estimation fPARMA}. All proofs are collected in Section~\ref{sec:Proofs}. Finally, Section~\ref{Sec: Conclusion} concludes.

\section{Preliminaries}\label{Section: Preliminaries}

\subsection{Notation}

The additive identity is denoted by $0$, and the identity map by $\mathbb{I}$. On a Cartesian product space $V^n$, $n \in \mathbb{N}$, the inner product of $x = (x_1, \dots, x_n)^{\!\top}$ and $y = (y_1, \dots, y_n)^{\top} \in V^n$ is defined by $\langle x, y \rangle = \sum_{i=1}^n \langle x_i, y_i \rangle$, and the norm by $\|x\| = (\sum_{i=1}^n \|x_i\|^2)^{1/2}$, where $V$ is a linear space equipped with inner product $\langle \cdot, \cdot \rangle$ and norm $\|\cdot\|$. Let $(\mathcal{H}, \langle \cdot, \cdot \rangle)$ and $(\mathcal{H}_\star, \langle \cdot, \cdot \rangle_\star)$ be Hilbert spaces. We denote by $\mathcal{L}_{\mathcal{H}, \mathcal{H}_\star}$ and $\mathcal{S}_{\mathcal{H}, \mathcal{H}_\star}$ the spaces of bounded linear operators and Hilbert-Schmidt (H-S) operators from $\mathcal{H}$ to $\mathcal{H}_\star$, respectively. These spaces are equipped with the operator norm $\|\cdot\|_{\mathcal{L}}$ and the H-S inner product $\langle \cdot, \cdot \rangle_{\mathcal{S}}$ with corresponding norm $\|\cdot\|_{\mathcal{S}}$. For $\mathcal{T} \in \{\mathcal{L}, \mathcal{S}\}$, we write $\mathcal{T}_{\mathcal{H}} \coloneqq \mathcal{T}_{\mathcal{H}, \mathcal{H}}$. For $x \in \mathcal{H}$ and $y \in \mathcal{H}_\star$, we define the \emph{tensor product operator} by $x \otimes y \coloneqq \langle x, \cdot \rangle y:\mathcal{H} \to \mathcal{H}_\star$. All random variables are defined on a common probability space $(\Omega, \mathfrak{A}, \Prob)$. For $p \in [1, \infty)$, let $L^p_{\mathcal{H}} = L^p_{\mathcal{H}}(\Omega, \mathfrak{A}, \Prob)$ denote the space of $\mathcal{H}$-valued random variables $X$ satisfying $\mathbb{E}\|X\|^p < \infty$. The \emph{cross-covariance operator} of $X \in L^2_{\mathcal{H}}$ and $Y \in L^2_{\mathcal{H}_\star}$ is defined by
\[
    \mathscr{C}_{\!X,Y}
    \coloneqq
    \mathbb{E}\bigl[(X - \mathbb{E}X) \otimes (Y - \mathbb{E}Y)\bigr],
\]
and the \emph{covariance operator} of $X$ by $\mathscr{C}_{\!X} \coloneqq \mathscr{C}_{\!X,X},$ with expectations in the Bochner sense.

\subsection{Cyclostationarity, periodic correlatedness and WN}\label{Subsec stat period}

Throughout, let $(\mathcal{H}, \langle \cdot, \cdot \rangle)$ be a separable Hilbert space. A process $(X_k) \subset \mathcal{H}$ is called \emph{strictly stationary} if
\[
    \big(X_{t_1}, X_{t_2}, \dots, X_{t_n}) \stackrel{d}{=} \big(X_{t_1+h}, X_{t_2+h} \dots, X_{t_n+h}\big),
    \quad \text{for all } h, t_1, \dots, t_n, n,
\]
and \emph{weakly stationary} (also referred to as \emph{second-order} or \emph{wide-sense stationary}) if
\[
    (X_k) \subset L^2_{\mathcal{H}}\,, \quad
    \Exp(X_k)=\mu \in \mathcal{H}
    \,\text{ for all } k, \quad
    \mathscr{C}_{X_k, X_\ell} = \mathscr{C}_{X_{k+h}, X_{\ell+h}}\,\text{ for all } h, k, \ell .
\] 
Moreover, a weakly stationary process $(X_k)$ is a \emph{weak white noise} (WWN) if
\[
    \Exp(X_k)=0 \text{ and }\Exp\!\|X_k\|^2>0 \,\text{ for all } k, \quad\mathscr{C}_{X_k, X_\ell}=0 \,\text{ for } k \neq \ell,
\]
and an i.i.d.~WWN is called \emph{strong white noise} (SWN). When no distinction between SWN and WWN is required, we simply write \emph{white noise} (WN).

A sequence $(a_k)_{k \in \mathbb{Z}}$ is called \emph{periodic} if $a_k = a_{k+T}$ for some $T \in \mathbb{N}$ and all $k$. The smallest such $T$ is called \emph{period}, and the sequence is said to be \emph{$T$-periodic}. In this case, $(a_k)$ is completely determined by $\{a_1, \dots, a_T\}$. For time series, $T$-periodicity is understood almost surely. More general concepts, such as approximate periodicity \citep{KoneMonsan2023} and poly-periodicity \citep{Gardner1993}, are not considered here. We now extend the notions of stationarity and white noise to the periodic setting, giving rise to the concepts commonly referred to as \emph{cyclostationarity} and \emph{periodic correlatedness}, the latter also known as \emph{wide-sense cyclostationarity} \citep[cf.][]{GardnerNapolitanoPaura2006}.

\begin{Def} A process $(X_k)$ is called \emph{cyclostationary} of period $T\in\mathbb{N}$ ($T$-CS), if
\[
    \big(X_{t_1}, X_{t_2}, \dots, X_{t_n}\big) \stackrel{d}{=} \big(X_{t_1+hT}, X_{t_2+hT}, \dots, X_{t_n+hT}\big)
    \quad\text{for all } h, t_1, \dots, t_n, n,
\]
and \emph{periodically correlated} ($T$-PC) of period $T\in\mathbb{N},$ if 
\[
    (X_k) \subset L^2_{\mathcal{H}}\,, \quad
    \Exp(X_k) = \Exp(X_{k+T})\,\text{ for all } k, \quad
    \mathscr{C}_{X_k, X_\ell} = \mathscr{C}_{X_{k+T}, X_{\ell+T}}\,\text{ for all } k, \ell.
\]
\end{Def}

\begin{Def} A $T$-PC  $(X_k)$ is a \emph{weak white noise} ($T$-PCWWN) if 
\[
    \Exp(X_k)=0 \text{ and }\Exp\!\|X_k\|^2>0 \,\text{ for all } k, \quad \mathscr{C}_{X_k, X_\ell} = 0
    \,\text{ for } k \text{ and } \ell \text{ in different cycles},
\]
that is, $k \in \{iT+1, \dots, iT+T\}$ and $\ell \in \{jT+1, \dots, jT+T\}$ with $i \neq j$. If, in addition, $(X_k)$ is independent and $T$-CS, it is called \emph{strong white noise} ($T$-PCSWN).
\end{Def}

\noindent When the distinction is immaterial, we write $T$-PCWN to refer to either the weak or strong version of the white noise. For any multiple $T'$ of $T$, $T$-CS ($T$-PC) implies $T'$-CS ($T'$-PC), and every $T$-PCWN is also a $T'$-PCWN. For $T=1$, the standard notions of stationarity and white noise are recovered. Moreover,
\[
    Y_k = \big(X_{(k-1)T+1}, X_{(k-1)T+2}, \dots, X_{(k-1)T+T}\big)^\top
\]
establishes a one-to-one correspondence between cyclostationary (periodically  correlated) processes of period $T$ in $\mathcal{H}$ and strictly (weakly) stationary processes in $\mathcal{H}^T$.

\begin{Ex} Let $(X_k)$ be a sequence of independent random variables with $X_{2k} \sim \mathcal{N}(\mu_X, \sigma_X)$ and $X_{2k+1} \sim \mathcal{N}(\mu_Y, \sigma_Y)$ for all $k \in \mathbb{Z}$. If $(\mu_X, \sigma_X) \neq (\mu_Y, \sigma_Y)$, then $(X_k)$ is a $2$-PCSWN.
\end{Ex}

\section{Functional PARMA Model}\label{Section Model}

Recall that a process $(X_k)_{k \in \mathbb{Z}}$ with values in a separable Hilbert space $\mathcal{H}$ with inner product $\langle\cdot, \cdot\rangle$ is called a \emph{functional autoregressive moving average process} with orders $p,q \in \mathbb{N}$ (fARMA$(p, q)$), if
\begin{align*}
  X_k - \mu = \sum_{i=1}^{p} \phi_i (X_{k-i} - \mu) + \sum_{j=1}^{q} \psi_j (\varepsilon_{k-j}) + \varepsilon_k, \quad k \in \mathbb{Z},
\end{align*}
where $\mu \in \mathcal{H}$, $(\varepsilon_k)_{k \in \mathbb{Z}} \subset \mathcal{H}$ denotes a WN, and $\phi_i, \psi_j \in \mathcal{L}_{\mathcal{H}}$ are bounded linear operators with $\phi_p \neq 0$ and $\psi_q \neq 0$. Moreover, for $q=0$, $(X_k) \subset \mathcal{H}$ is called a \emph{functional autoregressive process with order $p\in\mathbb{N}$} (fAR$(p)$). 

\subsection{Model formulation}

Let $T \in \mathbb{N}$. Further, let $(\varepsilon_k)_{k \in \mathbb{Z}} \subset \mathcal{H}$ be a $T$-PCWN, let $\mathsf{p}, \mathsf{q} \colon \mathbb{Z} \to \mathbb{N}$ be $T$-periodic functions, and assume that $(\mu_k)_{k \in \mathbb{Z}} \subset \mathcal{H}$ and $(\phi_{i,k})_{k \in \mathbb{Z}}, (\psi_{j,k})_{k \in \mathbb{Z}} \subset \mathcal{L}_{\mathcal{H}}$ are $T$-periodic sequences for each $i, j \in \mathbb{N}$. When the sequence has a period $T$, the individual index within the period is called season. The sequence is then indexed by $k = T\ell + s$ with cycle $\ell$ and season $s$.

In the following, we establish a functional ARMA-type process for which the mean, orders, and operators depend on the season $\ell = 1, \dots, T$, that is $\mu$ is replaced by $\mu_{\ell}$, the orders $p = \mathsf{p}(\ell)$ and $q = \mathsf{q}(\ell)$ vary with $\ell$, and the operators are given by $\phi_{i} = \phi_{i,\ell}$ and $\psi_{j} = \psi_{j,\ell}$, with $\phi_{\mathsf{p}(\ell),\ell} \neq 0$ and $\psi_{\mathsf{q}(\ell),\ell} \neq 0$ for each $\ell \in \{1, \dots, T\}.$ This condition ensures that the maximal orders $\mathsf{p}(\ell)$ and $\mathsf{q}(\ell)$ are well-defined. If for some $\ell$ either $\phi_{\mathsf{p}(\ell),\ell} = 0$ or $\psi_{\mathsf{q}(\ell),\ell} = 0$ hold, the orders $\mathsf{p}(\ell)$ and $\mathsf{q}(\ell)$ are reduced accordingly. Inspired by \cite{Troutman1979} for real-valued periodic autoregressive processes, we assume that 
\[
    T > M_T \coloneqq \max\Big\{\mathsf{p}(1),\, \mathsf{q}(1),\, \mathsf{p}(2),\, \mathsf{q}(2),\, \dots,\, \mathsf{p}(T),\, \mathsf{q}(T)\Big\},
\]
and that $(X_k)$ is centered. These assumptions involve no loss of generality by defining:
\[
    p = \max\Big\{\mathsf{p}(1),\,\mathsf{p}(2), \,\dots, \,\mathsf{p}(T)\Big\}, \quad q = \max\Big\{\mathsf{q}(1),\,\mathsf{q}(2), \,\dots, \,\mathsf{q}(T)\Big\},
\]
and setting
\begin{alignat*}{2}
    \phi_{i,\ell} &= 0, \quad &&\mathsf{p}(\ell) < i \leq p,\\
    \psi_{j,\ell} &= 0, \quad &&\mathsf{q}(\ell) < j \leq q.
\end{alignat*}
It is worth noting that if $T \le M_T,$ one can choose a multiple $T'$ of $T$ such that $T' > M_T,$ and set $\phi_{i,\ell} = \psi_{j,\ell} = 0$ for all $i > p$ and $j > q.$ With this convention, we can now state the following definition.

\begin{Def}[fPARMA]\label{Definition des fPARMA models - equiv version} Let $p, q, T \in \mathbb{N}$ with $T > \max(p, q)$, let $(\varepsilon_k)_{k \in \mathbb{Z}} \subset \mathcal{H}$ be a $T$-PCWWN, and let $(\phi_{i,k})_{k \in \mathbb{Z}}$, $(\psi_{j,k})_{k \in \mathbb{Z}} \subset \mathcal{L}_{\mathcal{H}}$ be $T$-periodic sequences for each $i, j \in \mathbb{N}$. Then, $(X_k)_{k \in \mathbb{Z}} \subset \mathcal{H}$ is called \emph{functional periodic ARMA process of period $T$ and orders $p$ and $q$} \emph{(fPARMA$(T, p, q)$)} if
\begin{align}\label{Defintion fPARMA process - equiv version}
    X_k = \sum_{i=1}^{p} \phi_{i,k}(X_{k-i}) + \sum_{j=1}^{q} \psi_{j,k}(\varepsilon_{k-j}) + \varepsilon_k, \quad k \in \mathbb{Z},
\end{align}
where $\phi_{p,k} \neq 0$ and $\psi_{q,k} \neq 0$ for some $k \in \{1, \dots, T\}$. Moreover, for $q = 0$, $(X_k)$ is called a \emph{functional periodic AR process} of period $T$ and order $p$ (fPAR$(T, p)$).
\end{Def}

This natural extension of functional ARMA model in general, separable Hilbert spaces indeed allows to handle periodicity. This model can also be extended to Banach spaces, though with some additional technical complexity. The key structural properties from the Hilbert space setting remain valid but require reformulation; in particular, inner products in the definition of tensorial products must be replaced by more general functionals; see \citet[Section 1.4]{Bosq2000}. It should also be noted that, due to the one-to-one correspondence between $T$-stationary and $T$-dimensional stationary processes, one may alternatively concatenate the seasons to a full cycle and regard these as successive observations from a stationary process which leads to a simpler modeling and estimation framework. However, it relies on complete weekly cycles, obscures day-specific dynamics, and restricts forecasting to entire cycles rather than finer short-term horizons. In the following, we give a specific example for a fPARMA process.

\begin{Ex}\label{simple Example}Let $(X_k)_{k \in \mathbb{Z}}$ be a time series representing the daily electricity consumption (in kWh) of private households in a given region. Since consumption patterns typically vary across weekdays, it is natural to model $(X_k)$ as a functional time series in $\mathcal{H} = L^2[0,1]$ of period $T = 7$, corresponding to the seven days of the week. For indices of the form $k = 7\ell + s$ with season $s \in \{1, \dots, 7\}$ and cycle $\ell \in \mathbb{Z}$, the observation $X_k$ then corresponds to the $s$th weekday of week $\ell$ (e.g., Monday to Sunday). We consider the centered process $(X_k - \mu_k)$, where $\mu_k = \mu_{k+7} = \mathbb{E}(X_k).$ A suitable model for such data is an fPARMA process of period $T = 7$ and orders $p, q < 7$. For illustration, we take $p = 3$ and $q = 1$, yielding
\begin{align*}
    X_k - \mu_k = \sum_{i=1}^3 \phi_{i,k}(X_{k-i} - \mu_{k-i}) + \psi_{1,k}(\varepsilon_{k-1}) + \varepsilon_k, \quad k \in \mathbb{Z},
\end{align*}
where $\phi_{i,k}, \psi_{1,k} \in \mathcal{L}_{\mathcal{H}}$ are 7-periodic in $k$, i.e., $\phi_{i,k} = \phi_{i,k+7}$ and $\psi_{1,k} = \psi_{1,k+7}$ for all $k$, and $(\varepsilon_k)_{k \in \mathbb{Z}}$ is a 7-PCWWN. It is convenient to assume that all operators are of integral form, i.e., $\gamma \colon \mathcal{H} \to \mathcal{H}$ with $\gamma(x)(t) = \int_0^1 g(s,t)x(s)\,\mathrm{d}s$ for each $x \in \mathcal{H}$ and intra-day time $t \in [0,1],$ where the kernel $g \colon [0,1]^2 \to \mathbb{R}$ is square-integrable. This model permits weekday-specific dynamics: for instance, Monday consumption ($k = 7\ell + 1$) may depend on the previous three days, including the weekend, with $\phi_{3,k} \neq 0$ for such $k$, whereas for other weekdays $s \in \{2, \dots, 7\}$, only the first lag may be relevant, i.e., $\phi_{1,s} \neq 0$ and $\phi_{2,s} = \phi_{3,s} = 0$. A lag-one moving average component is incorporated via $\psi_{1,k} \neq 0$ for all $k$. For further functional data analyses of electricity consumption, see \cite{AnderssonJostein2010, fontana2019functional, Malloretal2018}. 
\end{Ex}

\subsection{Multivariate representation}\label{Subsection representations}

To formulate the conditions under which the fPARMA equation \eqref{Defintion fPARMA process - equiv version} admits stationary solutions, we employ block operator matrices acting on Cartesian product spaces of the underlying Hilbert space. These matrices capture the recursive structure of the process and naturally give rise to multivariate functional representations. A comprehensive reference for the theory and structure of block operator matrices --- particularly in connection with spectral properties and \emph{companion operators} --- is \cite{Tretter2008}.

In our setting, we work with $\mathcal{H}^T$-valued processes defined by
\begin{gather}
    \boldsymbol{X}_{k} \coloneqq \scalebox{1}{$\begin{pmatrix}
	   X_{k-T+1}\\
	   X_{k-T+2}\\
       \vdots\\
	   X_{k}
	\end{pmatrix}
    $}, \qquad 
     \boldsymbol{\pi}_k \coloneqq \scalebox{1}{$\begin{pmatrix}
	   0\\
	   \vdots\\
       0\\
	   \varepsilon_{k}
	\end{pmatrix}
    $}, \qquad
    \boldsymbol{\varepsilon}_k \coloneqq  \scalebox{1}{$\begin{pmatrix}
	   \varepsilon_{k-T+1}\\
	   \varepsilon_{k-T+2}\\
       \vdots\\
	   \varepsilon_{k}
	\end{pmatrix}
    $}, \qquad k \in \mathbb{Z}.
    \label{eq:Def vector-valued processes}
\end{gather}

\noindent Setting $\phi_{i,k} = \psi_{j,k} = 0$ for $i>p,\, j>q$, and $\psi_{0,k} = \mathbb{I},$ we introduce the block operator matrices $\Phi_{k}, \Psi_{k} \in \mathcal{L}_{\mathcal{H}^T}$ as
\begin{align}\label{matrices mult repr}
    \Phi_{k} \coloneqq 
  \begin{pmatrix}
    0 & \mathbb{I} & 0 & \cdots & 0 \\
		\vdots & \ddots & \ddots & \ddots & \vdots\\
		\vdots & \ddots & \ddots & \ddots & 0\\
        0 & \cdots & \cdots & 0 & \mathbb{I}\\
        0 & \phi_{T-1,k} & \cdots & \phi_{2,k} & \phi_{1,k}
  \end{pmatrix}, 
  \qquad
  \Psi_{k} \coloneqq 
  \begin{pmatrix}
    0 & \cdots & \cdots & 0 \\
    \vdots & \cdots & \cdots & \vdots\\
    0 & \cdots & \cdots & 0\\
		\psi_{T-1,k} & \cdots & \psi_{1,k} & \psi_{0,k}
  \end{pmatrix},
  \quad k \in \mathbb{Z}.
\end{align}
Furthermore, we define the successive $i$-fold composition of the matrices $\Phi_{k}$ by
\begin{gather}\label{eq:mult operator-valued block matrix}
    \Phi_{i,k} \coloneqq \Phi_{k}\,\Phi_{k-1}\,\cdots\,\Phi_{k-(i-1)}, 
    \quad i\in\{1, 2, \dots,T\},\; k\in\mathbb{Z}, 
    \qquad 
    \Phi_{0,k} \coloneqq \mathbb{I}, 
    \quad k\in\mathbb{Z}.
\end{gather}

\noindent To establish sufficient conditions for cyclostationarity and periodic correlatedness,   and to support consistent estimation procedures, it is essential to specify how block operator matrices operate on elements of product spaces. For illustration, we demonstrate this in the case of $2 \times 2$ operator matrices.
\begin{align*}
    \scalebox{1}{$\begin{pmatrix}
        A_{11} & A_{12} \\
        A_{21} & A_{22}
    \end{pmatrix}
    $}
    \scalebox{1}{$\begin{pmatrix}
        x_1 \\
        x_2
    \end{pmatrix}
    $}
    &\coloneqq
    \scalebox{1}{$\begin{pmatrix}
        A_{11}(x_1) + A_{12}(x_2) \\
        A_{21}(x_1) + A_{22}(x_2)
    \end{pmatrix}
    $}, \\
    \scalebox{1}{$\begin{pmatrix}
        A_{11} & A_{12} \\
        A_{21} & A_{22}
    \end{pmatrix}
    $}
    \scalebox{1}{$\begin{pmatrix}
        B_{11} & B_{12} \\
        B_{21} & B_{22}
    \end{pmatrix}
    $}
    &\coloneqq
    \scalebox{1}{$\begin{pmatrix}
        A_{11}B_{11} + A_{12}B_{21} & A_{11}B_{12} + A_{12}B_{22} \\
        A_{21}B_{11} + A_{22}B_{21} & A_{21}B_{12} + A_{22}B_{22}
    \end{pmatrix}
    $},
\end{align*}

\vspace{1.5ex}

\noindent where $A_{ij}B_{k\ell}$ denotes the standard composition of bounded linear operators. In addition, establishing cyclostationarity and periodic correlatedness  requires the innovations to form PCWNs. For simplicity, we simultaneously present statements for both strict and weak stationarity and WNs, where the formulations in parentheses always refer to the weak versions.

\begin{As}\label{P-WN assumption} $(\varepsilon_k)_{k\in\mathbb{Z}} \subset \mathcal{H}$ is a $T\text{-PCSWN}$ $(T\text{-PCWWN})$, with $T \in \mathbb{N}$.
\end{As}

\begin{Prop}\label{Prop: WN feature transfer to the multivariate process} Let Assumption \ref{P-WN assumption} hold. Then, the following holds true.
\begin{itemize}  
    \item[\textnormal{(a)}] The process $(\boldsymbol{\rho}_{k})_{k\in\mathbb{Z}}\subset \mathcal{H}^T$ defined by
    \[
        \boldsymbol{\rho}_{k} \coloneqq \sum^T_{j=1}\,\Phi_{T-j,T}(\boldsymbol{\pi}_{\!(k-1)T+j}), \quad k \in\mathbb{Z},
    \]
    is a SWN (WWN), provided it holds $\boldsymbol{\rho}_{k} \neq 0$ almost surely.
    \item[\textnormal{(b)}] The process $(\boldsymbol{\varepsilon}'_{k})\subset \mathcal{H}^T,$ with $\boldsymbol{\varepsilon}'_{k} \coloneqq \boldsymbol{\varepsilon}_{kT},$ is a SWN (WWN). 
\end{itemize}
\end{Prop}

\subsubsection{fPAR processes} 

In analogy to the real-valued case considered in \cite{Troutman1979}, the multivariate formulation of an fPAR$(T,p)$ process $(X_k)$ can be expressed as an $\mathcal{H}^T$-valued fAR$(1)$ process, where the corresponding operator coefficients vary with the season. Specifically, we have
\begin{align*}
    \boldsymbol{X}_{k} = \Phi_{k}(\boldsymbol{X}_{k-1}) +  \boldsymbol{\pi}_k, \quad k\in\mathbb{Z},   
\end{align*}
where each $\Phi_k$ is an block operator matrix that encodes the seasonal structure of the process. Iterating this relation yields a representation over $T$ periods:
\begin{align*}
    \boldsymbol{X}_{k}
    &= \Phi_{T,k}(\boldsymbol{X}_{k-T}) + \Phi_{T-1,k}(\boldsymbol{\pi}_{k-(T-1)}) +\, \cdots \,+ \Phi_{0,k}( \boldsymbol{\pi}_k), \quad k\in\mathbb{Z},\allowdisplaybreaks
\end{align*}
where the coefficient matrices are defined in Eq.~\eqref{eq:mult operator-valued block matrix}. As the sequence  $(\Phi_{i,k})_k$ is $T$-periodic for every fixed $i$, the subsequence $(\boldsymbol{X}'_k) \subset \mathcal{H}^T$, defined by $\boldsymbol{X}'_k \coloneqq \boldsymbol{X}_{kT}$, evolves according to the relation
\begin{align}\label{eq:mult ar1}
    \boldsymbol{X}'_k = \boldsymbol{\Phi}(\boldsymbol{X}'_{k-1}) + \boldsymbol{\rho}_{k}, \quad k\in\mathbb{Z},
\end{align}
where $\boldsymbol{\Phi}$ is the block operator matrix given by the $T$-fold composition
\begin{align}\label{eq:def op.-val. prod matrix Phi}
    \boldsymbol{\Phi} \coloneqq \Phi_{T,T} = \Phi_{T}\,\Phi_{T-1}\,\cdots\,\Phi_{1}.    
\end{align}
Since $\boldsymbol{\Phi}$ is independent of the seasons, and because $(\boldsymbol{\rho}_k) \subset \mathcal{H}^T$ is a WWN by Proposition~\ref{Prop: WN feature transfer to the multivariate process}, the process $(\boldsymbol{X}'_k)$, which captures the full seasonal cycle within each observation, constitutes an $\mathcal{H}^T$-valued fAR$(1)$ process. 

\subsubsection{fPARMA processes} 

Similar to the situation in the fPAR setting, we derive a simpler multivariate representation of fPARMA processes. However, due to the more general structure of the fPARMA model, this representation includes an additional moving average component. Specifically, we have
\begin{align*}
    \boldsymbol{X}_{k} &= \Phi_{T,k}(\boldsymbol{X}_{k-T}) + \Phi_{T-1,k}\Psi_{k-(T-1)}(\boldsymbol{\varepsilon}_{k-(T-1)}) + \dots + \Phi_{0, k}\Psi_{k}(\boldsymbol{\varepsilon}_k), \quad k \in\mathbb{Z}.
\end{align*}

\noindent As in the fPAR case, we consider the $T$-subsampled process $(\boldsymbol{X}'_k) \subset \mathcal{H}^T$, defined by $\boldsymbol{X}'_k \coloneqq \boldsymbol{X}_{kT}$, along with the corresponding innovations $(\boldsymbol{\varepsilon}'_k)$. The periodicity of $(\Phi_{T,k})_k$ and $(\Psi_k)_k$, together with the structure of $(\boldsymbol{\varepsilon}'_k)$, leads to the
\begin{align}\label{periodic fARMA(B) is B^T fARMA(1,1)}
  \boldsymbol{X}'_k = \boldsymbol{\Phi}(\boldsymbol{X}'_{k-1}) + \boldsymbol{\Delta}_1(\boldsymbol{\varepsilon}'_{k-1}) + \boldsymbol{\Delta}_0(\boldsymbol{\varepsilon}'_{k}), \quad k\in\mathbb{Z},
\end{align}
where $\boldsymbol{\Phi}$ is the time-invariant operator from the fPAR case, and $\boldsymbol{\Delta}_0, \boldsymbol{\Delta}_1 \colon \mathcal{H}^T \to \mathcal{H}^T$ are operators that capture the aggregated moving average structure across the seasonal cycle. These are defined as
\begin{gather}\label{Def Delta_0 and Delta_1}
    \boldsymbol{\Delta}_0 \coloneqq \left(\,\sum_{\ell=0}^{T-j} \left(\Phi_{\ell,T} \Psi_{T-\ell}\right)_{i,j+\ell} \right)_{i,j=1}^T, \quad\;\,\boldsymbol{\Delta}_1 \coloneqq \left(\,\sum_{\ell=2}^{j} \left(\Phi_{T+1-\ell,T} \Psi_{\ell-1}\right)_{i,j+1-\ell} \right)_{i,j=1}^T,
\end{gather}
where each summand in $\boldsymbol{\Delta}_0$ and $\boldsymbol{\Delta}_1$ takes the form
\begin{alignat}{2}
    \left(\Phi_{\ell,T} \Psi_{T-\ell}\right)_{i,j+\ell} 
        &= \left(\Phi_{\ell,T}\right)_{i,T} \psi_{T-\ell-j, T-\ell}, 
        &&\quad 1 \leq j \leq T,~0 \leq \ell \leq T-j,\label{summands of Delta_0}\\
    \left(\Phi_{T+1-\ell, T} \Psi_{\ell-1}\right)_{i,j+1-\ell} 
        &= \left(\Phi_{T+1-\ell,T}\right)_{i,T} \psi_{T+\ell-j-1, \ell-1}, 
        &&\quad 2 \leq j \leq T,~2 \leq \ell \leq j,\label{summands of Delta_1}
\end{alignat}
and where the entries in the first column of $\boldsymbol{\Delta}_1$ are zero operators. Equation~\eqref{periodic fARMA(B) is B^T fARMA(1,1)} reveals that $(\boldsymbol{X}'_k)$ follows a multivariate fARMA$(1,1)$ structure, where, unlike the standard case, the coefficient $\boldsymbol{\Delta}_0$ need not be the identity map. By Proposition~\ref{Prop: WN feature transfer to the multivariate process}, $(\boldsymbol{\varepsilon}'_k)$ is a WWN, ensuring the representation is well-posed. Similar formulations with general operator $\boldsymbol{\Delta}_0$ also appear in \citet{Spangenberg2013}.

\section{Properties}\label{sec:properties}

In the following, we derive conditions for cyclostationarity and periodic correlatedness of period $T\in\mathbb{N},$ and discuss structural properties of our fPARMA process $(X_k)\subset\mathcal{H}.$ 

\subsection{Cyclostationarity and periodic correlatedness}

In order to state a result on cyclostationarity and periodic correlatedness of fPAR(MA) processes, we need the following

\begin{As}\label{Operator assumption f(T)ARMA} $\|\boldsymbol{\Phi}^{j_0}\|_{\mathcal{L}} < 1$ for some $j_0\in\mathbb{N}.$
\end{As}

\begin{Theo}\label{Stationarity of fPARMA processes} Let Assumptions \ref{P-WN assumption}--\ref{Operator assumption f(T)ARMA} hold. Then the fPARMA$(T,p,q)$ equation \eqref{Defintion fPARMA process - equiv version} admits a unique, $T$-CS (\,$T$-PC) causal solution $(X_k)\subset\mathcal{H}$, with
\[
    X_{kT+j} \coloneqq \boldsymbol{X}'^{\,(j)}_k, \quad j=1,\dots,T,\; k\in\mathbb{Z},
\]
where $\boldsymbol{X}'_k=\boldsymbol{X}_{kT}\in\mathcal{H}^T$ is defined as in \eqref{periodic fARMA(B) is B^T fARMA(1,1)} and $\boldsymbol{X}_{\ell}$ is defined in \eqref{eq:Def vector-valued processes}. Moreover, 
\begin{align}\label{lin process representation of multivariate efARMA(1,1)}
    \boldsymbol{X}'_k = \sum_{j=1}^\infty\,\boldsymbol{\Phi}^{j-1}(\boldsymbol{\Phi}\boldsymbol{\Delta}_0 + \boldsymbol{\Delta}_1)(\boldsymbol{\varepsilon}'_{k-j}) + \boldsymbol{\Delta}_0(\boldsymbol{\varepsilon}'_k), \quad k\in\mathbb{Z},
\end{align}
where the series converges in $L^2$ and almost surely.
\end{Theo}

\subsection{Structural properties}

In order to establish a result on finite moments, finite moment conditions of the corresponding innovations are required.

\begin{As}\label{Assumption: nu-th moments of errors} For $(\varepsilon_k)$ in Assumption \ref{P-WN assumption}, it holds $\sup_k\Exp\!\|\varepsilon_k\|^{\tau} < \infty$ for some $\tau\geq2.$
\end{As}

\begin{Prop}\label{Prop: Finite higher moments} Let Assumptions \ref{P-WN assumption}
--\ref{Assumption: nu-th moments of errors} hold. Then, $\sup_k\Exp\!\|X_k\|^{\tau}<\infty.$ 
\end{Prop}
 
\noindent In what follows, let $\mathscr{C}^h_{\!\boldsymbol{X}'} \coloneqq \mathscr{C}_{\!\boldsymbol{X}'_0, \boldsymbol{X}'_h}$ be the \emph{lag-$h$ covariance operators} of the $T$-dimensional stationary process $\boldsymbol{X}'=(\boldsymbol{X}'_k) = (\boldsymbol{X}_{kT})_k \subset \mathcal{H}^T\!,$ associated with the $T$-CS fPAR(MA) process, from which the lag-$h$ covariance operators of $\boldsymbol{X}$ for each season can be recovered.

\begin{Prop}\label{Covariance structure} Suppose Assumptions \ref{P-WN assumption}--\ref{Operator assumption f(T)ARMA} hold. Then, the following holds true.
\begin{itemize}
    \item[\textnormal{(a)}] Let $(X_k)$ be a  fPAR$(T,p)$ process. Then,  
    \begin{align*}
        \mathscr{C}^h_{\!\boldsymbol{X}'} = 
        \begin{cases}
            \,\boldsymbol{\Phi}^h\mathscr{C}_{\!\boldsymbol{X}'}, & \textrm{if } h \geq 0,\\[1ex]
            \,\mathscr{C}_{\!\boldsymbol{X}'}\boldsymbol{\Phi}^{\ast h}, & \textrm{if } h < 0.
        \end{cases}    
    \end{align*}
    Moreover, for the covariance operator $\mathscr{C}_{\!\boldsymbol{X}'}$, it holds
    \begin{align*}
        \mathscr{C}_{\!\boldsymbol{X}'} = \sum^\infty_{i=0}\,\boldsymbol{\Phi}^i\mathscr{C}_{\!\boldsymbol{\rho}}\boldsymbol{\Phi}^{\ast i},   
    \end{align*}
    where the covariance operator of the WWN $\boldsymbol{\rho} = (\boldsymbol{\rho}_k)$ satisfies
    \begin{align*}
        \mathscr{C}_{\!\boldsymbol{\rho}} = \sum^T_{i=1}\,\Phi_{T-i,T}\mathscr{C}_{\boldsymbol{\pi}_{\!i}}\Phi^\ast_{T-i,T}, \quad \text{with}\;\;
        \mathscr{C}_{\boldsymbol{\pi}_{\!i}} = \mathrm{diag}\big(0, \dots, 0, \mathscr{C}_{\varepsilon_i}\big).
    \end{align*}  

    \item[\textnormal{(b)}] Let $(X_k)$ be a fPARMA$(T,p,q)$ process. Then, 
        \begin{align}\label{Representation of lag-covariance operators ARMA}
            \mathscr{C}^h_{\!\boldsymbol{X}'} = 
            \begin{cases} 
                \,\boldsymbol{\Phi}^h\mathscr{C}_{\!\boldsymbol{X}'} + \boldsymbol{\Phi}^{h-1}\!\boldsymbol{\Delta}_1\mathscr{C}_{\boldsymbol{\varepsilon}'}\boldsymbol{\Delta}^\ast_0, &\textrm{if } h > 0, \\[-2ex]
                &\\
                \,\mathscr{C}_{\!\boldsymbol{X}'}\boldsymbol{\Phi}^{\ast h} + \boldsymbol{\Delta}_0\mathscr{C}_{\boldsymbol{\varepsilon}'}\boldsymbol{\Delta}^\ast_1\boldsymbol{\Phi}^{\ast h-1}, & \mbox{if } h < 0. 
            \end{cases}
        \end{align}  
    In addition, with $\boldsymbol{\tilde{\Phi}}_i\!\coloneqq \boldsymbol{\Phi}^{i-1}(\boldsymbol{\Phi}\boldsymbol{\Delta}_0 + \boldsymbol{\Delta}_1)$ for $i\in\mathbb{N}$ and $\boldsymbol{\tilde{\Phi}}_0\coloneqq\boldsymbol{\Delta}_0,$ it holds 
    \begin{gather}\label{Representation of covariance operator ARMA w.r.t. covariance operators of errors}   
        \mathscr{C}_{\!\boldsymbol{X}'} = \sum^\infty_{i=0}\boldsymbol{\tilde{\Phi}}_i\mathscr{C}_{\boldsymbol{\varepsilon}'}\boldsymbol{\tilde{\Phi}}^\ast_i,\allowdisplaybreaks
    \end{gather}
    where the covariance operator of the WWN $\boldsymbol{\varepsilon}' = (\boldsymbol{\varepsilon}'_k)$ has the form 
    \begin{align}\label{covariance op of epsilons}
        \mathscr{C}_{\boldsymbol{\varepsilon}'} = \mathrm{diag}\big(\mathscr{C}_{\varepsilon_1}, \mathscr{C}_{\varepsilon_2}, \dots, \mathscr{C}_{\varepsilon_T}\big).    
    \end{align}
\end{itemize}
\end{Prop}

\noindent Finally, we establish weak dependence in the sense of \cite{HoermannKokoszka2010}, who introduced the concept of \emph{$L^\tau$-$m$-approximability}. For $\tau \geq 1,$ a process $(Y_{k})_{k\in\mathbb{Z}} \subset L^\tau_{\mathcal{H}}$ is said to be $L^\tau$-$m$-approximable if it admits a \emph{Bernoulli shift representation}, that is 
\[
    Y_{k} = f(\varepsilon_{k}, \varepsilon_{k-1}, \ldots), \quad k\in\mathbb{Z},
\]
for some measurable function $f\colon S^{\mathbb{N}} \rightarrow \mathcal{H}$ and an i.i.d.\ process $(\varepsilon_{k})_{k\in\mathbb{Z}} \subset S$ on a measurable space $S$, such that
\[
    \sum_{m=1}^\infty\,\nu_\tau\!\left(Y_m - Y^{(m)}_m\right) < \infty.
\]
Here, $\nu_\tau(\cdot) =  (\Exp\!\|\cdot\|^\tau)^{1/\tau},$ and
\[
    Y^{(m)}_{k} = f\big(\varepsilon_{k}, \ldots, \varepsilon_{k-m+1}, \varepsilon^{(k)}_{k-m}, \varepsilon^{(k)}_{k-m-1}, \ldots\big),
\] 
where $(\varepsilon^{(n)}_{k})_k$ denote independent copies of $(\varepsilon_{k})$ for each $n$. This form of weak dependence permits consistent estimation of means, (lagged) (cross-)covariance operators, and eigenelements  \citep[see, e.g.,][]{HoermannKokoszka2010, Kuehnert2024}; and for $\tau \geq 2$, it further implies a central limit theorem \citep{HoermannKokoszka2010}.

\begin{Prop}\label{Prop: Weak dependence} Let Assumptions \ref{P-WN assumption}--\ref{Operator assumption f(T)ARMA} and \ref{Assumption: nu-th moments of errors}\,$(\tau)$ hold. Then the process $(X_k)$ is $L^\tau$-$m$-approximable with geometrically decaying approximation errors, i.e.,
\[
    \nu_\tau\!\left(X_m\!- X^{(m)}_m\right) \leq c \rho^m
\]
for all sufficiently large $m$ for some $c>0$ and $\rho\in(0,1)$.
\end{Prop}

\begin{Rem}For a rigorous treatment of limit theorems in the context of fPAR$(1)$ processes in Hilbert spaces, see \citet[Section~3]{SoltaniHashemi2011a}. Further, limit theorems for the related functional linear processes can be found in \citet[Section~7]{Bosq2000}.
\end{Rem}

\section{Estimation of the fPAR operators}\label{Section: Estimation fPAR}

This section develops estimators for the fPAR$(T,p)$ operators, under the assumption that the period $T$ is known, and where $p<T.$ 

\subsection{Estimation procedure}

We begin by estimating the block operator matrix $\boldsymbol{\Phi}$ in Eq.~\eqref{eq:def op.-val. prod matrix Phi}, which characterizes the underlying fAR process. From a sample $X_1, \dots, X_N$ of the fPAR process we obtain an estimator $\hat{\boldsymbol{\Phi}} = \hat{\boldsymbol{\Phi}}_{\!N}$, which we treat as given in the following and focus on subsequent steps. For details on the estimation of fAR operators, see \cite{Bosq2000, CaponeraPanaretos2022} \citep[see also][]{AueKuehnertRice2025}. Note that compactness is a crucial property when operators are estimated via finite-dimensional approximations. For simplicity, we assume the following.

\begin{As}\label{As: Consistent estimates} All fPAR operators $\phi_{i,j}$ are H-S (thus also $\boldsymbol{\Phi}$). 
\end{As}

In the following we extract estimators for the fPAR operators from the given matrix $\hat{\boldsymbol{\Phi}}$. Hereto, we use that the entries of $\boldsymbol{\Phi}$ are recursively defined by
\begin{gather*}
    \boldsymbol{\Phi}_{i,j} = \mathds{1}_{\{i+1, \dots, T\}}(j) \, \phi_{T+i-j,i} + \sum_{k=1}^{i-1}\,\phi_{k,i} \boldsymbol{\Phi}_{(T-i)-k,j}, \quad 1 \leq i, j \leq T.
\end{gather*}

\noindent The structure is illustrated by
\begin{align*}
    \resizebox{\textwidth}{!}{$
    \boldsymbol{\Phi} = 
\begin{pmatrix}
    0 
    & \phi_{T-1,1} 
    & \phi_{T-2,1} 
    & \phi_{T-3,1}
    & \cdots 
    & \phi_{1,1} \\
    0 
    & \phi_{1,2} \boldsymbol{\Phi}_{1,2} 
    & \phi_{T-1,2} + \phi_{1,2} \boldsymbol{\Phi}_{1,3} 
    & \phi_{T-2,2} + \phi_{1,2} \boldsymbol{\Phi}_{1,3}
    & \cdots 
    & \phi_{2,2} + \phi_{1,2} \boldsymbol{\Phi}_{1,T} \\
    0 
    & \sum_{k=1}^{2} \phi_{k,3} \boldsymbol{\Phi}_{3-k,2}
    & \sum_{k=1}^{2} \phi_{k,3} \boldsymbol{\Phi}_{3-k,3} 
    & \phi_{T-2,3} + \sum_{k=1}^{2} \phi_{k,3} \boldsymbol{\Phi}_{3-k,4}
    & \cdots 
    & \phi_{3,3} + \sum_{k=1}^{2} \phi_{k,3} \boldsymbol{\Phi}_{3-k,T}\\
    \vdots 
    & \vdots 
    & \ddots 
    & \ddots
    & \ddots
    & \vdots \\
    \vdots 
    & \sum_{k=1}^{P-2} \phi_{k,T-1} \boldsymbol{\Phi}_{(T-1)-k,2} 
    & \sum_{k=1}^{P-2} \phi_{k,T-1} \boldsymbol{\Phi}_{(T-1)-k,3}
    & \sum_{k=1}^{P-2} \phi_{k,T-1} \boldsymbol{\Phi}_{(T-1)-k,4}
    & \cdots 
    & \phi_{T-1,T-1} + \sum_{k=1}^{P-2} \phi_{k,T-1} \boldsymbol{\Phi}_{(T-1)-k,T} \\
    0 
    & \sum_{k=1}^{T-1} \phi_{k,T} \boldsymbol{\Phi}_{T-k,2} 
    & \sum_{k=1}^{T-1} \phi_{k,T} \boldsymbol{\Phi}_{T-k,3}
    & \sum_{k=1}^{T-1} \phi_{k,T} \boldsymbol{\Phi}_{T-k,4}
    & \cdots 
    & \sum_{k=1}^{T-1} \phi_{k,T} \boldsymbol{\Phi}_{T-k,T}
\end{pmatrix}
$}
\end{align*}

\vspace{1.5ex}

\noindent Inspecting the first row shows that the estimators for each $\phi_{j,1}$ can be directly recovered from $\hat{\boldsymbol{\Phi}}.$ Moreover, the entries in each $j$th row contain $j$ unknown parameters in the lower-triangular part of the remaining non-trivial part of the matrix:
\begin{gather}\label{Matrix repr for estimation}
    \boldsymbol{\phi}_m\boldsymbol{\Phi}_{[mm]} = \boldsymbol{\Phi}_{[m]}, \quad 2 \leq m \leq T,
\end{gather}
where 
\begin{align*}
    \boldsymbol{\phi}_m &\coloneqq \bigl(\phi_{m-1,m}, \phi_{m-2,m}, \dots, \phi_{1,m}\bigr), \\[0.3em]
    \boldsymbol{\Phi}_{[mm]} &\coloneqq \bigl( \boldsymbol{\Phi}_{i-1,j} \bigr)_{i,j=2}^m, \\[0.3em]
    \boldsymbol{\Phi}_{[m]} &\coloneqq \bigl(\boldsymbol{\Phi}_{m,2}, \boldsymbol{\Phi}_{m,3}, \dots, \boldsymbol{\Phi}_{m,m}\bigr).
\end{align*}

Since the full block operator matrix $\boldsymbol{\Phi}$ can be consistently estimated under Assumption~\ref{As: Consistent estimates}, we proceed by constructing estimators for the fPAR operators using empirical analogues. To address the inherent ill-posedness, we apply Tikhonov regularization~\citep{Tikhonov1943}; for a comprehensive review of regularization methods in statistics, see \cite{Bickeletal2006}. Another frequently used approach is based on Moore-Penrose pseudoinverses~\citep{moore1920,penrose1955}. Throughout, we use Tikhonov-regularized inverses which we define for an operator $A$ by
\[
    A^\dagger = A^\ast \big(AA^\ast + \vartheta_{\!N} \mathbb{I}\big)^{-1},
\]
where $\mathbb{I}$ is the identity map and $\vartheta_{\!N} \to 0$ a sequence of positive regularization parameters. Within this framework, each block operator coefficient vector $\boldsymbol{\phi}_m$ is identifiable from Eq.~\eqref{Matrix repr for estimation}, provided that the principal submatrices $\boldsymbol{\Phi}_{[mm]}$ of $\boldsymbol{\Phi}_{[TT]}$ have dense image, as ensured under 

\begin{As}\label{as:dense image}
  The operator $\boldsymbol{\Phi}_{[TT]}$ has dense image.
\end{As}

\noindent Given this assumption, and based on the structure implied by Eq.~\eqref{Matrix repr for estimation}, we estimate $\boldsymbol{\phi}_m$ via
\begin{align}\label{eq: estimate for op vectors fPAR}
    \hat{\boldsymbol{\phi}}_m \coloneqq \hat{\boldsymbol{\Phi}}_{[m]}\hat{\boldsymbol{\Phi}}_{[mm]}^{\dagger}\hat{\mathscr{P}}_m, \quad 2 \le m \leq T.
\end{align}
Here, $\hat{\boldsymbol{\Phi}}_{[mm]}$ and $\hat{\boldsymbol{\Phi}}_{[m]}$ are empirical counterparts of the corresponding theoretical blocks, and
\[
    \hat{\boldsymbol{\Phi}}_{[mm]}^{\dagger} \coloneqq \hat{\boldsymbol{\Phi}}_{[mm]}^{\ast} \left( \hat{\boldsymbol{\Phi}}_{[mm]} \hat{\boldsymbol{\Phi}}_{[mm]}^{\ast} + \vartheta^{[m]}_{\!N}\, \mathbb{I} \right)^{-1}
\]
is the Tikhonov-regularized inverse with positive parameters $\vartheta^{[m]}_{\!N} \to 0$ as $N \to \infty$. The operator $\hat{\mathscr{P}}_m$ projects onto the span of the eigenfunctions associated to the leading $K^{[m]} = K^{[m]}_{\!N}$ eigenvalues $\hat{\varphi}_{m,1} \ge \dots \ge \hat{\varphi}_{m,K^{[m]}}$ of the Gram operator $\hat{\boldsymbol{\Phi}}_{[mm]} \hat{\boldsymbol{\Phi}}_{[mm]}^{\ast}$. \noindent The notation makes explicit that both the regularization parameter $\vartheta^{[m]}_{\!N}$ and the truncation level $K^{[m]}_{\!N}$ may be chosen adaptively for each $m$, for instance based on the decay behavior of the empirical eigenvalues \citep[cf.][]{KuehnertRiceAue2025}. Since only finitely many such operators $\hat{\boldsymbol{\Phi}}_{[mm]}$ occur in practice, one may also choose these sequences uniformly across $m$, albeit potentially at the expense of slower convergence rates.

Finally, combining the estimates from the first row of $\hat{\boldsymbol{\Phi}}$ with those obtained via \eqref{eq: estimate for op vectors fPAR}, we construct estimators for all fPAR operators. The estimation procedure is summarized as follows:

\begin{algorithm}\caption{fPAR Operator Estimation}\label{alg:AR}
    \begin{algorithmic}[1]
        \Require Estimator $\hat{\boldsymbol{\Phi}}$; period $T$; tuning parameters \(\vartheta^{[m]}_N \), \(K^{[m]}_N \) for \( 2 \leq m < T \)
        \Ensure Estimators \( \hat{\phi}_{k,\ell} \) for \( 1 \leq k < T \), \( 1 \leq \ell \leq T \)
        \For{$k = 1$ to $T-1$}
            \State $\hat{\phi}_{k,1} \gets \hat{\Phi}_{1,\,T+1-k}$
        \EndFor
        \For{$\ell = 2$ to $T$}
            \State Compute $\hat{\boldsymbol{\phi}}_\ell \gets \hat{\boldsymbol{\Phi}}_{[\ell]} \hat{\boldsymbol{\Phi}}_{[\ell\ell]}^{\dagger}\hat{\mathscr{P}}_{\ell}$
            
            \For{$k = 1$ to $\ell-1$}
                \State $\hat{\phi}_{k,\ell} \gets \hat{\boldsymbol{\phi}}^{(\ell-k)}_\ell$
            \EndFor
            
            \For{$k = \ell$ to $T-1$}
                \State $\hat{\phi}_{k,\ell} \gets \hat{\boldsymbol{\Phi}}_{\ell,T+\ell-k} - \sum_{m=1}^{k-1} \hat{\phi}_{m,\ell} \, \hat{\boldsymbol{\Phi}}_{\ell-m,T+\ell-k}$
            \EndFor
        \EndFor
    \end{algorithmic}
\end{algorithm}

\begin{Ex}\label{ex:Algorithm}
Let $T = 3$ and $p = 2.$ Further, suppose we are given the block operator matrix $\boldsymbol{\Phi}$ and its estimator $\hat{\boldsymbol{\Phi}},$ defined by
\[
    \boldsymbol{\Phi} = 
    \scalebox{1}{$\begin{pmatrix}
        0 & \boldsymbol{\Phi}_{1,2} & \boldsymbol{\Phi}_{1,3} \\
        0 & \boldsymbol{\Phi}_{2,2} & \boldsymbol{\Phi}_{2,3} \\
        0 & \boldsymbol{\Phi}_{3,2} & \boldsymbol{\Phi}_{3,3}
    \end{pmatrix}
    $}
    \qquad\text{resp.}\qquad
    \hat{\boldsymbol{\Phi}} = 
    \scalebox{1}{$\begin{pmatrix}
        0 & \hat{\boldsymbol{\Phi}}_{1,2} & \hat{\boldsymbol{\Phi}}_{1,3} \\
        0 & \hat{\boldsymbol{\Phi}}_{2,2} & \hat{\boldsymbol{\Phi}}_{2,3} \\
        0 & \hat{\boldsymbol{\Phi}}_{3,2} & \hat{\boldsymbol{\Phi}}_{3,3}
    \end{pmatrix}.
    $}
\]
Then, the required subvectors and submatrices in Algorithm~\ref{alg:AR} have the form: 
    \begin{alignat*}{2}
       \hat{\boldsymbol{\Phi}}_{[2]} &= \hat{\boldsymbol{\Phi}}_{2,2}, \quad &&\hat{\boldsymbol{\Phi}}_{[22]} = \hat{\boldsymbol{\Phi}}_{1,2},\\
       \hat{\boldsymbol{\Phi}}_{[3]} &= \left(\hat{\boldsymbol{\Phi}}_{3,2}, \hat{\boldsymbol{\Phi}}_{3,3}\right), \quad &&\hat{\boldsymbol{\Phi}}_{[33]} =     \scalebox{1}{$\begin{pmatrix}
          \hat{\boldsymbol{\Phi}}_{1,2} & \hat{\boldsymbol{\Phi}}_{1,3} \\
          \hat{\boldsymbol{\Phi}}_{2,2} & \hat{\boldsymbol{\Phi}}_{2,3} \\
       \end{pmatrix}
       $}\,.
    \end{alignat*}

\noindent Furthermore, by defining the Tikhonov-regularized inverses $\hat{\boldsymbol{\Phi}}_{[\ell\ell]}^{\dagger}$ and the projection operators $\hat{\mathscr{P}}_{\ell}$ of the Gram operators $\hat{\boldsymbol{\Phi}}_{[\ell\ell]} \hat{\boldsymbol{\Phi}}_{[\ell\ell]}^{\ast}$ for $\ell\in\{2,3\},$ applying Algorithm~\ref{alg:AR} leads to the estimators 
    \begin{alignat*}{2}
        \hat{\phi}_{1,1} &= \hat{\boldsymbol{\Phi}}_{1,3}, ~~ &&\hat{\phi}_{2,1}= \hat{\boldsymbol{\Phi}}_{1,2},\\
        \hat{\phi}_{1,2} &= \hat{\boldsymbol{\phi}}_2^{(1)}\! = \!\left(\hat{\boldsymbol{\Phi}}_{[2]} \hat{\boldsymbol{\Phi}}_{[22]}^{\dagger}\hat{\mathscr{P}}_{2}\right)^{\!(1)}, \qquad &&\hat{\phi}_{2,2} = \hat{\boldsymbol{\Phi}}_{2,3} - \hat{\phi}_{1,2} \, \hat{\boldsymbol{\Phi}}_{1,3},\\
        \hat{\phi}_{1,3} &= \hat{\boldsymbol{\phi}}_3^{(2)}\! = \!\left(\hat{\boldsymbol{\Phi}}_{[3]} \hat{\boldsymbol{\Phi}}_{[33]}^{\dagger}\hat{\mathscr{P}}_{3}\right)^{\!(2)}, \qquad &&
        \hat{\phi}_{2,3} = \hat{\boldsymbol{\phi}}_3^{(1)} = \!\left(\hat{\boldsymbol{\Phi}}_{[3]} \hat{\boldsymbol{\Phi}}_{[33]}^{\dagger}\hat{\mathscr{P}}_{3}\right)^{\!(1)}\,.
    \end{alignat*}
\end{Ex}

\subsection{Consistency} 

Unless stated otherwise, all limits are understood as $N \to \infty$. To derive consistency for the full fPAR operators, we assume the following:

\begin{As}\label{As: Conv Ps-inv and Ps-id} For $N\to\infty$ holds 
    \begin{gather}\label{cons op matrix}
        \big\|\hat{\boldsymbol{\Phi}} - \boldsymbol{\Phi}\big\|_\mathcal{S} = o_{\Prob}(1),\\
        \max_{2\leq \ell \leq T}\bigg\{\,\Big\|\hat{\boldsymbol{\Phi}}^{\dagger}_{[\ell\ell]}\hat{\mathscr{P}}_{\ell} -\boldsymbol{\Phi}^{\dagger}_{[\ell\ell]} \mathscr{P}_{\ell}\Big\|_{\mathcal{L}} ~,~ \Big\|\left(\boldsymbol{\Phi}^{\ddagger}_{[\ell\ell]} \mathscr{P}_{\ell} - \mathbb{I}\right)\boldsymbol{\phi}_{\ell}\Big\|_{\mathcal{S}}\,\bigg\} = o_{\Prob}(1),\label{abstract cond}
  \end{gather}
where $\hat{\mathscr{P}}_{\ell}$ and $\mathscr{P}_{\ell}$ are the projection operators on the spans of the eigenfunctions associated to the $K=K^{[\ell]}_{\!N}$ leading eigenvalues of the Gram operators $\hat{\boldsymbol{\Phi}}_{[\ell\ell]}\hat{\boldsymbol{\Phi}}^{\ast}_{[\ell\ell]}$ and $\boldsymbol{\Phi}_{[\ell\ell]} \boldsymbol{\Phi}^{\ast}_{[\ell\ell]},$ respectively. 
\end{As}

The seemingly abstract condition~\eqref{abstract cond} is satisfied in simple settings. The first term in the maximum, involving consistent estimation of projected Tikhonov-regularized inverses, holds under mild assumptions using inequalities from \cite{Reimherr2015}. The second term, requiring that the operators $\boldsymbol{\phi}_{\ell}$ are well-approximated by finite-dimensional projections, is fulfilled, for instance, under a Sobolev-type condition used for estimating operators of invertible linear and AR(MA) processes in \cite{KuehnertRiceAue2025}. Both cases rely on consistent estimation of (lagged) (cross-)covariance operators under weak dependence \citep[see, e.g.,][]{Kuehnert2024}.

\begin{Theo}\label{Theo:estimation errors fPAR} Let Assumptions \ref{As: Consistent estimates}--\ref{As: Conv Ps-inv and Ps-id} hold. Then, as $N\to \infty,$
\begin{alignat}{2}
    &\max_{1 \le k < T}\,\bigl\|\hat{\phi}_{k,1} - \phi_{k,1}\bigr\|_{\mathcal{S}}
    = \mathcal{O}_{\Prob}\!\left(\bigl\|\hat{\boldsymbol{\Phi}} - \boldsymbol{\Phi}\bigr\|_{\mathcal{S}}\right),  \qquad && \label{rate 1st ln}
\end{alignat}
and for each $2\leq \ell \leq T,$ we have
\begin{alignat}{2}
    &\max_{1 \le k < T}\,\bigl\|\hat{\phi}_{k,\ell} - \phi_{k,\ell}\bigr\|_{\mathcal{S}}\notag\\
    &\quad = \mathcal{O}_{\Prob}\Biggl(
      \max\Biggl\{
        \varphi_{\!\ell, K^{[\ell]}_{N}}^{-1/2}\,
        \bigl\|\hat{\boldsymbol{\Phi}} - \boldsymbol{\Phi}\bigr\|_{\mathcal{S}} ~,~ 
        \Bigl\|
          \hat{\boldsymbol{\Phi}}^{\dagger}_{[\ell\ell]} \hat{\mathscr{P}}_{\ell}
          - \boldsymbol{\Phi}^{\dagger}_{[\ell\ell]} \mathscr{P}_{\ell}
        \Bigr\|_{\mathcal{L}} ~,~         \Bigl\|
          \Bigl(
            \boldsymbol{\Phi}^{\ddagger}_{[\ell\ell]} \mathscr{P}_{\ell}
            - \mathbb{I}
          \Bigr)\boldsymbol{\phi}_{\ell}
        \Bigr\|_{\mathcal{S}}
      \Biggr\}
    \Biggr).
    \label{rate others}
\end{alignat}
\end{Theo}

We illustrate explicit consistency rates in the setting of Example \ref{ex:Algorithm}.

\begin{Ex}\label{main example - AR} Recall Example \ref{ex:Algorithm}, where $T=3$ and $p=2.$ Assume
\begin{align*}
    \phi_{i,j} = c_{ij}\phi, \quad 1 \leq i \leq 2,~ 1\leq j \leq 3,
\end{align*}
where $\phi$ is a self-adjoint, positive definite H-S operator, and suppose $c_{12}=c_{23}=0$ and $c_{ij} \neq 0$ otherwise. By compactness and self-adjointness, $\phi$ admits the spectral decomposition
\begin{align}\label{op phi}
    \phi = \sum^\infty_{j=1}\,\varphi_j(e_j\otimes e_j),
\end{align}
where $\varphi_1 \geq \varphi_2 \geq \cdots > 0$ are square-summable eigenvalues with eigenfunctions $e_1, e_2, \dots\,.$ Further, for some $\beta > 0$ assume the Sobolev-type condition
\begin{align}\label{sob con phi}
    \sum^\infty_{j=1}\,\varphi^2_j(1+j^{2\beta}) < \infty.
\end{align}
By the definition of $\boldsymbol{\Phi}$ in \eqref{eq:def op.-val. prod matrix Phi} and the assumptions above, we obtain
\begin{align*}
  \boldsymbol{\Phi} &=  
  \begin{pmatrix}
    0 & c_{21}\phi & c_{11}\phi\\
    0 & c_{12}c_{21}\phi^2 & c_{22}\phi + c_{11}c_{12}\phi^2\\    
    0 & c_{12}c_{13}c_{21}\phi^3 + c_{21}c_{23}\phi^2 
      & c_{13}\phi(c_{22}\phi + c_{11}c_{12}\phi^2) + c_{11}c_{23}\phi^2
  \end{pmatrix}\\
  &= 
  \begin{pmatrix}
    0 & c_{21} & c_{11}\\
    0 & 0      & c_{22} \\    
    0 & 0      & c_{13}c_{22}\phi
  \end{pmatrix}\phi\,.
\end{align*}
For the submatrix $\boldsymbol{\Phi}_{[33]},$ we find, using self-adjointness of $\phi$,
\[
    \boldsymbol{\Phi}^\ast_{[33]} =  
    \begin{pmatrix}
        c_{21}\phi & c_{11}\phi\\
        0 & c_{22}\phi
    \end{pmatrix}^{\!\ast} = 
    \begin{pmatrix}
        c_{21} & 0\\
        c_{11} & c_{22}
    \end{pmatrix}\phi.
\]
This matrix is injective since $\phi$ is, and because we assumed $c_{21},\,c_{22}\neq 0.$ Further, due to the fact that the image of an operator lies dense if and only if its adjoint is injective, Assumption \ref{as:dense image} holds true.

Now, for $k=1,2,3,$ let $E_{i,k}\in\mathcal{H}^3$ be the vector with $e_i$ in the $k$th component and zeros elsewhere. Using $\langle e_i, e_j\rangle = \delta_{ij},$ where $\delta_{ij}$ denotes the \emph{Kronecker delta} (i.e.~$\delta_{ij} = 1$ for $i=j$ and $\delta_{ij}=0$ otherwise), together with \eqref{sob con phi}, it follows that
\begin{align*}
    &\sum^\infty_{i=1}\sum^\infty_{j=1}\sum^3_{k=1}\sum^3_{\ell=1}\, \big\langle\boldsymbol{\Phi}(E_{i,k}),  E_{j,\ell}\big\rangle^2_\mathcal{S}\,(1+i^{2\beta})\\
    &\quad = \sum^\infty_{i=1}\sum^\infty_{j=1}\sum^3_{k=1}\sum^3_{\ell=1}\,\varphi^2_i\left\langle\big(c_{21}\delta_{k2} + c_{11}\delta_{k3},~c_{22}\delta_{k3},~c_{13}c_{22}\varphi_i\delta_{k3} \big)^{\top}e_i,  E_{j,\ell}\right\rangle^2_\mathcal{S}\,(1+i^{2\beta})\\
    &\quad = \sum^\infty_{i=1}\,\varphi^2_i\left[c^2_{21} + c^2_{22}(2 + c^2_{13}\varphi^2_i)\right](1+i^{2\beta}) < \infty.
\end{align*}

\smallskip

\noindent Moreover, since $\boldsymbol{\Phi}$ is an fAR(1) operator (cf. \eqref{eq:mult ar1}), by using a Yule-Walker-type estimator $\hat{\boldsymbol{\Phi}}$ for $\boldsymbol{\Phi}$, for a suitable sequence $K=K_{\!N}\to\infty,$ we have \citep[see][Corollary 5.1]{KuehnertRiceAue2025}
\[
    \big\|\hat{\boldsymbol{\Phi}} - \boldsymbol{\Phi}\big\|_\mathcal{S} = \mathcal{O}_{\Prob}(K^{-\beta}) =  o_{\Prob}(1).
\]

\vspace{1.5ex}

\noindent Further, assuming $K=K^{[1]}_N = K^{[2]}_N$ for simplicity, and letting $\Lambda_K$ denote the $K$th reciprocal spectral gap of $\mathscr{C}_{\!\boldsymbol{X}'}$, for $N\to \infty$ it holds
\begin{align*}
    \max\bigg\{\,\Big\|\hat{\boldsymbol{\Phi}}^{\dagger}_{[22]}\hat{\mathscr{P}}_2 -\boldsymbol{\Phi}^{\dagger}_{[22]} \mathscr{P}_2\Big\|_{\mathcal{L}}~, ~\Big\|\hat{\boldsymbol{\Phi}}^{\dagger}_{[33]}\hat{\mathscr{P}}_3 -\boldsymbol{\Phi}^{\dagger}_{[33]} \mathscr{P}_3\Big\|_{\mathcal{L}}\,\bigg\} = \mathcal{O}_{\Prob}\big(\Lambda_KN^{-1/2}\big),
\end{align*}
and due to $\boldsymbol{\Phi}_{[22]} = c_{21}\phi$ and $\boldsymbol{\Phi}_{[33]}$ involving $\phi,$ one also has
\begin{align*}
    \max\bigg\{\,\Big\|\Big(\boldsymbol{\Phi}^{\ddagger}_{[22]} \mathscr{P}_2 - \mathbb{I}\Big)\boldsymbol{\phi}_2\Big\|_{\mathcal{S}}~,~\Big\|\Big(\boldsymbol{\Phi}^{\ddagger}_{[33]} \mathscr{P}_3 - \mathbb{I}\Big)\boldsymbol{\phi}_3\Big\|_{\mathcal{S}}\,\bigg\} = \mathcal{O}\big(K^{-\beta}\big).
\end{align*}
\noindent Altogether, since the asymptotic behavior of the eigenvalues $\varphi_j$ of $\phi$ matches (up to multiplicative constants) that of the eigenvalues $\varphi_{\ell,j}$ of $\boldsymbol{\Phi}_{[\ell\ell]} \boldsymbol{\Phi}^\ast_{[\ell\ell]}$, Theorem \ref{Theo:estimation errors fPAR} yields
\begin{align*}
    \max_{1\leq k < 3}\big\|\hat{\phi}_{k,1} - \phi_{k,1}\big\|_\mathcal{S} &= \mathcal{O}_{\Prob}\big(K^{-\beta}\big),\\[0.5ex]
    \max_{2\leq \ell \leq 3}\,\max_{1\leq k < 3}\big\|\hat{\phi}_{k,\ell} - \phi_{k,\ell}\big\|_\mathcal{S} &= \mathcal{O}_{\Prob}\left(\,\max\left\{\varphi^{-1/2}_{\!K}K^{-\beta},\,\Lambda_KN^{-1/2}\right\}\right).
\end{align*}

\vspace{1.5ex}

Finally, we derive an explicit rate in terms of $N$ alone, which requires additional assumptions. Assume $\varphi_j \sim j^{-a},$ $a>0,$ so that \eqref{sob con phi} holds with $\beta = a-1/2 - c$ for small $c>0.$ It remains to determine the rate of the reciprocal spectral gaps $\Lambda_j$ of $\mathscr{C}_{\!\boldsymbol{X}'}$ and admissible sequences $K=K_{\!N}\to\infty.$ Notice that, by Proposition \ref{Prop: WN feature transfer to the multivariate process}~(a), the covariance operator $\mathscr{C}_{\!\boldsymbol{X}'}$ satisfies
\begin{align*}
    \mathscr{C}_{\!\boldsymbol{X}'} = \sum^\infty_{i=0}\,\boldsymbol{\Phi}^i\mathscr{C}_{\!\boldsymbol{\rho}}\boldsymbol{\Phi}^{\ast i}\!,   
\end{align*}
where the covariance operator of the WWN $(\rho_k),$ using $T=3$ and the matrices in \eqref{eq:mult operator-valued block matrix}, satisfies
\begin{align*}
    \mathscr{C}_{\!\boldsymbol{\rho}} = \Phi_3\big(\Phi_2\mathscr{C}_{\boldsymbol{\pi}_{\!1}}\Phi_2^\ast + \mathscr{C}_{\boldsymbol{\pi}_{\!2}}\big)\Phi_3^\ast + \mathscr{C}_{\boldsymbol{\pi}_{\!3}},\quad 
    \mathscr{C}_{\boldsymbol{\pi}_{\!i}}=\mathrm{diag}\big(0,0,\mathscr{C}_{\varepsilon_i}\big),~i=1,2,3.
\end{align*}
For simplicity, assume $\mathscr{C}_{\varepsilon_i} = d_i\mathscr{C}_{\varepsilon}$, $i=1,2,3$, with positive constants $d_1,d_2,d_3$ and a covariance operator $\mathscr{C}_{\varepsilon}$, where 
\begin{align*}
    \mathscr{C}_{\varepsilon} = \sum^\infty_{j=1}\,a_j(e_j\otimes e_j),
\end{align*}
with $e_j$ matching the eigenfunctions of $\phi$ in \eqref{op phi}. By the definition of the operator-valued matrices $\Phi_2, \Phi_3$ in \eqref{matrices mult repr}, and since $\phi_{2,3}$ and $\phi_{1,2}$ vanish, and $\phi$ commutes with $\mathscr{C}_{\varepsilon},$ we obtain 
\begin{align*}
    \mathscr{C}_{\!\boldsymbol{\rho}} 
    &= 
    \begin{pmatrix}
        d_1 & 0 & 0\\
        0   & d_2 & d_2c_{13}\phi\\    
        0   & d_2c_{13}\phi & d_2c^2_{13}\phi^2 + d_3
    \end{pmatrix}\!\mathscr{C}_{\varepsilon}\,.
\end{align*}
The decay of the eigenvalues $\lambda_j$ of $\mathscr{C}_{\!\boldsymbol{X}'}$, expressed as sums depending only on $\phi$ and $\mathscr{C}_{\varepsilon}$ with matching eigenfunctions, coincides with that of $\mathscr{C}_{\!\boldsymbol{\rho}},$ which in turn matches the decay of $a_j$ (up to multiplicative constants) in $\mathscr{C}_{\varepsilon}.$ Under the assumption $a_j\sim j^{-2}$ as $j\to\infty,$ it holds $\Lambda_j = (\lambda_j-\lambda_{j-1})^{-1}\sim j^3.$ Theorem 3.1 in \cite{KuehnertRiceAue2025} requires $K^{\beta+1/2}\Lambda^2_K = \mathcal{O}(N^{1/2}),$ which, with $\beta = a - 1/2 - c >0,$ holds for $K\sim N^{1/2(6+a-c)}.$ Then, for sufficiently small $c>0,$ 
\[
    \max_{1\leq k < 3}\big\|\hat{\phi}_{k,1} - \phi_{k,1}\big\|_\mathcal{S} = \mathcal{O}_{\Prob}\left(N^{-\frac{2a-1-2c}{4a+24-4c)}}\right) \approx \mathcal{O}_{\Prob}\left(N^{-\frac{2a-1}{4a+24}}\right),
\]
and, since $\varphi_j \sim j^{-a},$ for $c>0$ sufficiently small, it holds that
\begin{align*}
    \max_{2\leq \ell \leq 3}\,\max_{1\leq k < 3}\big\|\hat{\phi}_{k,\ell} - \phi_{k,\ell}\big\|_\mathcal{S} &= \mathcal{O}_{\Prob}\left(\max\big\{N^{-\frac{a-1-2c}{4a+24-4c}},~N^{-\frac{a-c}{2a+12-2c}}\big\}\right)\\
    &\approx \mathcal{O}_{\Prob}\left(\max\big\{N^{-\frac{a-1}{4a+24}},~N^{-\frac{a}{2a+12}}\big\}\right)\\
    &= \mathcal{O}_{\Prob}\left(N^{-\frac{a-1}{4a+24}}\right)\,.
\end{align*}
Then, for instance, for $a=8,$ we have
    \begin{align*}
        \max_{1\leq k < 3}\big\|\hat{\phi}_{k,1} - \phi_{k,1}\big\|_\mathcal{S} \approx \mathcal{O}_{\Prob}\left(N^{-\frac{15}{56}}\right), \qquad \max_{2\leq \ell \leq 3}\,\max_{1\leq k < 3}\big\|\hat{\phi}_{k,\ell} - \phi_{k,\ell}\big\|_\mathcal{S} = \mathcal{O}_{\Prob}\left(N^{-\frac{1}{8}}\right)\,.
    \end{align*}
\end{Ex}

\smallskip

\begin{Rem} 
\mbox{}\\[-2ex]
    \begin{itemize}
        \item[\textnormal{(a)}] In suitable settings, the asymptotic rate in Eq.~\eqref{rate 1st ln} can approach the parametric rate $N^{-1/2}$ \citep[cf.][Example 5.1]{KuehnertRiceAue2025}. In general, larger $\beta>0$ and more slowly decaying eigenvalues $\lambda_j$ of $\mathscr{C}_{\!\boldsymbol{X}'}$ --- hence more slowly growing reciprocal gaps $\Lambda_j$ --- improve the rates in Theorem \ref{Theo:estimation errors fPAR} (cf.~Example \ref{main example - AR}).

        \item[\textnormal{(b)}] In practice, consistency results for finite-dimensional fPAR approximations are essential \citep[see][in the context of fAR(MA)]{KuehnertRiceAue2025}. If the fPAR operators admit such representations, consistency holds for $\phi_{i,1}$, $i \in \{1, \dots, T-1\}$, in the first row of $\boldsymbol{\Phi}$ in Eq.~\eqref{eq:def op.-val. prod matrix Phi}. In our approach, however, consistent estimation of the remaining fPAR operators requires $\boldsymbol{\Phi}_{[TT]}$ to have a dense image, inherited by $\boldsymbol{\Phi}_{[mm]}$, contradicting the finite-dimensional assumption and thus limiting these results.

        \item[\textnormal{(c)}] A weak convergence result for estimation errors toward a non-trivial limit is unavailable for the full operators in the multivariate fAR model underlying the fPAR operators, as shown in Theorem 3.2 of \cite{Mas2007}. Under certain technical conditions, however, Theorem 3.1 in the same reference establishes asymptotic normality for prediction errors at specific points. In functional linear regression, \cite{KuttaDierickxDette2022} derived a pivotal test statistic for the slope operator by restricting it to a suitable smoothness class. With additional assumptions, a similar result could extend to our setting.

        \item[\textnormal{(d)}] A simplification of Assumption~\ref{As: Conv Ps-inv and Ps-id} is feasible in exceptional cases. Although the eigenvalues of the principal submatrices $\boldsymbol{\Phi}_{[mm]}$ interlace those of $\boldsymbol{\Phi}_{[TT]}$ by Cauchy’s interlacing theorem this property does not necessarily extend to $\boldsymbol{\Phi}_{[mm]}\boldsymbol{\Phi}^{\ast}_{[mm]}$ and $\boldsymbol{\Phi}_{[TT]} \boldsymbol{\Phi}^{\ast}_{[TT]}$.
    \end{itemize}
\end{Rem} 

\section{Estimation in fPARMA models}\label{Section: Estimation fPARMA}

In this section, we study parameter estimation for the fPARMA model under a restricted assumption, which allows for a tractable and transparent analysis of the underlying estimation mechanism. For the case of PARMA models in the univariate setting, see \citet{adams1995,sarnaglia2015periodic,Vecchia1985} and in the multivariate setting, see \citet{AknoucheHamdi2009}.

Specifically, we consider an fPARMA$(2,1,1)$ model and assume that the operator $\boldsymbol{\Delta}_0$ in Eq.~\eqref{Def Delta_0 and Delta_1}---appearing in the representation of $(\boldsymbol{X}'_k)\subset\mathcal{H}^2$ in Eq.~\eqref{periodic fARMA(B) is B^T fARMA(1,1)}---equals the identity map. From \eqref{Def Delta_0 and Delta_1}--\eqref{summands of Delta_0} and the definitions of the involved operators, we obtain
\[
    \boldsymbol{\Delta}_0
    = \left(\,\sum_{\ell=0}^{2-j} (\Phi_{\ell,2})_{i,2}\,
      \psi_{2-\ell-j,\,2-\ell}\right)_{\!i,j=1}^{\!2}
    =
    \begin{pmatrix}
        \mathbb{I} & 0 \\
        \psi_{1,2} + \phi_{1,2} & \mathbb{I}
    \end{pmatrix},
\]
so that $\boldsymbol{\Delta}_0=\mathbb{I}$ holds if and only if $\psi_{1,2}=-\phi_{1,2}$. While this restriction is clearly strong and typically unverifiable in practice, it yields a setting in which the structure of the estimation problem becomes particularly transparent. Under this assumption, the process $(\boldsymbol{X}'_k)$ reduces to a standard fARMA$(1,1)$ model, for which estimation procedures are available; see, for example, \citet{kuenzer:2024} and \citet{KuehnertRiceAue2025}. We emphasize that even within this simplified framework, the general case cannot be recovered by a mere redefinition of the innovations $\tilde{\boldsymbol{\varepsilon}}'_k = \boldsymbol{\Delta}_0(\boldsymbol{\varepsilon}'_k)$. Such a transformation would require modifying all innovation-related terms by the (pseudo-)inverse of the unknown operator $\boldsymbol{\Delta}_0$, thereby substantially increasing the technical complexity of the estimation procedure.

We now discuss estimation of the fPARMA$(2,1,1)$ operators in this simplified setting. As in the fPAR case, we assume that estimators $\hat{\boldsymbol{\Phi}}=\hat{\boldsymbol{\Phi}}_{\!N}$ and $\hat{\boldsymbol{\Delta}}_1=(\hat{\boldsymbol{\Delta}}_1)_{\!N}$ for $\boldsymbol{\Phi}$ and $\boldsymbol{\Delta}_1$ are available, based on a sample $X_1,\dots,X_N$ from the fPARMA process. The fPAR estimators $\hat{\phi}_{1,j}$, $1\le j\le 2$, are extracted from $\hat{\boldsymbol{\Phi}}$ as described in Section~\ref{Section: Estimation fPAR}, while the fpMA estimators $\hat{\psi}_{1,j}$, $1\le j\le 2$, are derived analogously from $\hat{\boldsymbol{\Delta}}_1$. Since $\psi_{1,2}=-\phi_{1,2}$, we set
\[
    \hat{\psi}_{1,2} \coloneqq -\hat{\phi}_{1,2},
\]
so that only $\hat{\psi}_{1,1}$ remains to be estimated. For $T=2$, \eqref{Def Delta_0 and Delta_1} and \eqref{summands of Delta_1} yield
\[
    \boldsymbol{\Delta}_1
    =
    \begin{pmatrix}
        0 & (\Phi_{1,2})_{1,2}\,\psi_{1,1} \\
        0 & (\Phi_{1,2})_{2,2}\,\psi_{1,1}
    \end{pmatrix}
    =
    \begin{pmatrix}
        0 & \psi_{1,1} \\
        0 & \phi_{1,2}\psi_{1,1}
    \end{pmatrix},
\]
which motivates the estimator
\[
    \hat{\psi}_{1,1} \coloneqq (\hat{\boldsymbol{\Delta}}_1)_{1,2}.
\]

\vspace{1ex}

\begin{Cor}\label{theo:cons fPARMA} Let $(X_k)$ be a fPARMA$(2,1,1)$ process, and let $\boldsymbol{\Delta}_0 = \mathbb{I}.$ Further, let the conditions of Theorem \ref{Theo:estimation errors fPAR} for $T=2$ hold, and let $\psi_{1,1}, \psi_{1,2}$ be H-S. Then, as $N\to \infty,$
\begin{alignat*}{2}
    &\big\|\hat{\phi}_{1,1} - \phi_{1,1}\big\|_\mathcal{S} = \mathcal{O}_{\Prob}\!\left(\bigl\|\hat{\boldsymbol{\Phi}} - \boldsymbol{\Phi}\bigr\|_{\mathcal{S}}\right), \quad   \big\|\hat{\psi}_{1,1} - \psi_{1,1}\big\|_\mathcal{S} = \mathcal{O}_{\Prob}\left(\big\|\hat{\boldsymbol{\Delta}}_1 - \boldsymbol{\Delta}_1\big\|_{\mathcal{S}}\right), \label{rate 1st ln}\\[1ex]
    &\big\|\hat{\phi}_{1,2} - \phi_{1,2}\big\|_\mathcal{S} = \big\|\hat{\psi}_{1,2} - \psi_{1,2}\big\|_\mathcal{S}\notag\\
    &\; = \mathcal{O}_{\Prob}\left(\max\left\{\varphi^{-1/2}_{\!2, K^{[2]}_{\!N}}\,\big\|\hat{\boldsymbol{\Phi}} - \boldsymbol{\Phi}\big\|_{\mathcal{S}} ~,~ \Big\|\hat{\boldsymbol{\Phi}}^{\dagger}_{[22]}\hat{\mathscr{P}}_2 -\boldsymbol{\Phi}^{\dagger}_{[22]} \mathscr{P}_2\Big\|_{\mathcal{L}} ~,~ \Big\|\left(\boldsymbol{\Phi}^{\ddagger}_{[22]} \mathscr{P}_2 - \mathbb{I}\right)\boldsymbol{\phi}_2\Big\|_{\mathcal{S}}\right\}\right).
\end{alignat*} 
\end{Cor}

\section{Proofs}\label{sec:Proofs}

This section provides the proofs of all our theoretical results. 

\smallskip 

\begin{proof}[\textbf{Proof of Proposition \ref{Prop: WN feature transfer to the multivariate process}}]
We show that the processes in parts (a)–(b) are weak white noises (WWNs). The fact that they are also strong white noises (SWNs), provided the underlying error processes satisfy the corresponding assumptions, follows immediately.

\vspace{2ex}

\noindent (a)\, Suppose
\[
    \boldsymbol{\rho}_{k} = \sum^T_{j=1}\,\Phi_{T-j,T}\big(\boldsymbol{\pi}_{(k-1)T+j}\big)\neq 0 ~\text{ a.s.}, \quad \text{for each } k, 
\] 
where $\boldsymbol{\pi}_k = (0, \dots, 0, \varepsilon_k)^\top.$ Since $(\varepsilon_k)$ is a $T$-PCWWN by Assumption~\ref{P-WN assumption}, we obtain
\begin{align}
    0 < \Exp\!\|\boldsymbol{\rho}_{k}\|^2 
    &= \Exp\!\bigg\|\sum^T_{j=1}\Big((\Phi_{T-j,T})_{1,T}(\varepsilon_{(k-1)T+j}), \dots, (\Phi_{T-j,T})_{T,T}(\varepsilon_{(k-1)T+j})\Big)^{\!\top}\,\bigg\|^2\notag\allowdisplaybreaks\\ 
    &\leq T\,\sum^T_{j=1}\sum^T_{\ell=1}\big\|(\Phi_{T-j,T})_{\ell,T}\big\|^2_{\mathcal{L}}\Exp\!\big\|\varepsilon_{(k-1)T+j}\big\|^2\notag\allowdisplaybreaks\\ 
    &= T^2\!\max_{1\leq j,\ell \leq T}\big\|(\Phi_{T-j,T})_{\ell,T}\big\|^2_{\mathcal{L}}\,\sum^T_{j=1}\,\Exp\!\|\varepsilon_{j}\|^2 < \infty.\notag
\end{align}
Moreover, $\Exp(\varepsilon_{k}) = 0$ for all $k$ implies $\Exp(\boldsymbol{\rho}_{k}) = 0$ for all $k.$ For any $k,\ell\in\mathbb{Z}$, we further obtain
\begin{align*}
    \mathscr{C}_{\!\boldsymbol{\rho}_{k}, \boldsymbol{\rho}_{\ell}} 
    &= \sum^T_{i=1}\sum^T_{j=1}\,\Phi_{T-j,T}\mathscr{C}_{\boldsymbol{\pi}_{\!(k-1)T+i}, \boldsymbol{\pi}_{\!(\ell-1)T+j}}\Phi^\ast_{T-i,T}\notag\\
    &= \delta_{k\ell}\sum^T_{i=1}\,\Phi_{T-i,T}\mathscr{C}_{\boldsymbol{\pi}_{\!i}}\Phi^\ast_{T-i,T},
\end{align*}
where the second equality follows from the fact that $(\boldsymbol{\pi}_k)$ is a $T$-PCWWN. As a consequence, $(\boldsymbol{\rho}_{k})$ is a WWN.

\vspace{2ex}

\noindent (b)\, Since $(\boldsymbol{\varepsilon}_{k})$ is a $T$-PCWWN, it follows that
\begin{align*}
    \Exp(\boldsymbol{\varepsilon}'_{k}) = 0
    ~\text{ and }~  \Exp\!\|\boldsymbol{\varepsilon}'_{k}\|^2
    = \sum^T_{j=1}\,\Exp\!\|\varepsilon_{ j}\|^2\, \in \,(0,\infty) \quad \text{for all } k.    
\end{align*}
Moreover, the relations $\mathscr{C}_{\varepsilon_{k}} = \mathscr{C}_{\varepsilon_{k+T}}$ for all $k$ and $\mathscr{C}_{\varepsilon_{k}, \varepsilon_{\ell}} = 0$ for $k \neq \ell$ imply 
\[
    \mathscr{C}_{\boldsymbol{\varepsilon}'_{k}, \boldsymbol{\varepsilon}'_{\ell}}
    = \sum^T_{j=1}\,\Big(
    \Exp\!\big\langle\varepsilon_{(k-1)T + j}, \cdot\big\rangle\varepsilon_{(\ell-1)T + 1},
    \dots,
    \Exp\!\big\langle\varepsilon_{(k-1)T + j}, \cdot\big\rangle\varepsilon_{(\ell-1)T + T}
    \Big)^{\!\top}, \quad \text{for each } k, \ell,
\]
thus $\mathscr{C}_{\boldsymbol{\varepsilon}'_{k}} = \mathscr{C}_{\boldsymbol{\varepsilon}'_{\ell}}$ for all $k,\ell$, and $\mathscr{C}_{\boldsymbol{\varepsilon}'_{k}, \boldsymbol{\varepsilon}'_{\ell}} = 0$ for $k \neq \ell$. Therefore, $(\boldsymbol{\varepsilon}'_{k})$ is indeed a WWN.
\end{proof}

\smallskip

\begin{proof}[\textbf{Proof of Theorem  \ref{Stationarity of fPARMA processes}}] Weak stationarity and almost sure uniqueness of the $\mathcal{H}^T$-valued fARMA$(1,1)$ process $(\boldsymbol{X}'_k)$ in \eqref{periodic fARMA(B) is B^T fARMA(1,1)}, driven by a weak (strong) white noise $(\boldsymbol{\varepsilon}'_k)$ as in Proposition~\ref{Prop: WN feature transfer to the multivariate process}, follow directly from \citet[Theorem~3.4]{KlepschKlueppelbergWei2017} in the case $\boldsymbol{\Delta}_0=\mathbb{I}$. A careful inspection of their arguments shows that the result extends without difficulty to the case $\boldsymbol{\Delta}_0\neq\mathbb{I}$, since the proofs rely only on boundedness and summability conditions of the involved operators.

The linear process representation \eqref{lin process representation of multivariate efARMA(1,1)} follows by the same reasoning, with only minor adaptations to the present framework. In particular, the corresponding series converges both in $L^2$ and almost surely.

Moreover, by causality, the process $(\boldsymbol{X}'_k)$ is strictly stationary whenever $(\boldsymbol{\varepsilon}'_k)$ is a strong white noise; see
\citet[][Theorem~36.4]{Billingsley1995}. Finally, exploiting the one-to-one correspondence between cyclostationarity (periodic correlatedness) of period $T$ of processes, and $T$-dimensional strictly (weakly) stationary time series, the process  $(X_k)$ defined by $X_{kT+j} \coloneqq \boldsymbol{X}'^{\,(j)}_k$ admits a unique, causal solution that is cyclostationary (periodically correlated) of period $T.$
\end{proof}

\smallskip

\begin{proof}[\textbf{Proof of Proposition \ref{Prop: Finite higher moments}}] From the linear process representation \eqref{lin process representation of multivariate efARMA(1,1)} and the sub-multiplicativity of the operator norm, it follows that
\[
    \Exp\!\|\boldsymbol{X}'_k\|^{\tau} \leq 2^{\tau-1}\!\left[\,\big\|\boldsymbol{\Phi}\boldsymbol{\Delta}_0 + \boldsymbol{\Delta}_1\big\|^{\tau}_{\mathcal{L}}\,\Exp\!\left(\,\sum^\infty_{j=1}\,\|\boldsymbol{\Phi}^{j-1}\|_{\mathcal{L}}\|\boldsymbol{\varepsilon}'_{k-j}\|\right)^{\!\tau} + \,\|\boldsymbol{\Delta}_0\|^{\tau}_{\mathcal{L}}\Exp\!\|\boldsymbol{\varepsilon}'_{k}\|^{\tau}\,\right]\!, ~~\tau \geq 2.
\]
Moreover, according to the definition of $\boldsymbol{\varepsilon}'_{k}$ and Assumption~\ref{Assumption: nu-th moments of errors}\,$(\tau)$, it holds that
\[
    \Exp\!\|\boldsymbol{\varepsilon}'_{k}\|^{\tau} = \Exp\!\left(\,\sum^{T-1}_{i=0}\|\varepsilon_{k-i}\|^2 \right)^{\!\tau/2} \leq T^{\,\tau/2}\sup_{k}\Exp\!\|\varepsilon_{k}\|^{\tau} < \infty.
\]
Further, by using Minkowski’s inequality, the monotone convergence theorem, and Assumption~\ref{Operator assumption f(T)ARMA}—which is equivalent to $\|\boldsymbol{\Phi}^j\|_{\mathcal{L}} < ab^j$ for all $j \in \mathbb{N}$ and some $a>0$, $0<b<1$ \citep[][Lemma~3.1]{Bosq2000}-we obtain
\begin{align*}
    \Exp\!\left(\,\sum^\infty_{j=1}\,\big\|\boldsymbol{\Phi}^{j-1}\big\|_{\mathcal{L}}\|\boldsymbol{\varepsilon}'_{k-j}\|\right)^{\tau}
    &\leq a^\tau T^{\,\tau/2}\sup_k\Exp\!\|\varepsilon_{k}\|^{\tau}
    \left(\,\sum^\infty_{j=0}b^j\right)^{\!\tau} < \infty.
\end{align*}
Combining the above bounds then yields $\sup_k\Exp\!\|\boldsymbol{X}'_k\|^{\tau}<\infty$, and hence, as claimed, $\sup_k\Exp\!\|X_{k}\|^{\tau}<\infty.$
\end{proof}

\smallskip

\begin{proof}[\textbf{Proof of Proposition \ref{Covariance structure}}] \noindent (a)\, The result follows directly from the definitions of $\boldsymbol{\rho}_k$ and $\boldsymbol{\pi}_k$ \citep[cf.][Proposition~3.3]{Bosq2000}.

\vspace{2ex}

\noindent (b)\, By \eqref{periodic fARMA(B) is B^T fARMA(1,1)}, the definition of lag-$h$ covariance operators, the fact that $\boldsymbol{X}'_k$ is independent of $\boldsymbol{\varepsilon}'_{\ell}$ for $k<\ell$, and since $(\boldsymbol{\varepsilon}'_k)$ is a WWN, we obtain for $h>0$
\begin{align*}
\mathscr{C}^h_{\!\boldsymbol{X}'}
    &\,= \boldsymbol{\Phi}\mathscr{C}^{h-1}_{\!\boldsymbol{X}'} + \boldsymbol{\Delta}_1\!\Exp\langle\boldsymbol{X}'_{\!0}, \cdot\rangle\boldsymbol{\varepsilon}'_{h-1} + \boldsymbol{\Delta}_0\!\Exp\langle\boldsymbol{X}'_{\!0}, \cdot\rangle\boldsymbol{\varepsilon}'_{h}\notag\\[-0.5ex]
    &\quad\!\!\vdots\notag\\
    &\,= \boldsymbol{\Phi}^h\mathscr{C}_{\!\boldsymbol{X}'} + \boldsymbol{\Phi}^{h-1}\!\boldsymbol{\Delta}_1\!\Exp\!\big\langle\boldsymbol{\Phi}(\boldsymbol{X}'_{\!-1}) + \boldsymbol{\Delta}_1(\boldsymbol{\varepsilon}'_{-1}) + \boldsymbol{\Delta}_0(\boldsymbol{\varepsilon}'_{0}), \cdot \big\rangle\boldsymbol{\varepsilon}'_{0}.
\end{align*}
This proves \eqref{Representation of lag-covariance operators ARMA} for $h>0$. The representation for $h<0$ follows from the identity $(\mathscr{C}^h_{\!\boldsymbol{X}'})^\ast=\mathscr{C}^{-h}_{\!\boldsymbol{X}'}$, $h\in\mathbb{Z}$.

Moreover, by using \eqref{lin process representation of multivariate efARMA(1,1)}, weak stationarity, the definition of $\boldsymbol{\tilde{\Phi}}^j$, $j\in\mathbb{N}$, and the fact that $(\boldsymbol{\varepsilon}'_k)$ is a WWN, we obtain
\begin{align*}
    \mathscr{C}_{\!\boldsymbol{X}'}
    &= \sum^\infty_{i=1}\sum^\infty_{j=1}\delta_{ij}\!\Exp\langle \boldsymbol{\tilde{\Phi}}_i(\boldsymbol{\varepsilon}'_{\!-i}), \cdot\rangle\boldsymbol{\tilde{\Phi}}_j(\boldsymbol{\varepsilon}'_{\!-j}) + \Exp\langle\boldsymbol{\Delta}_0(\boldsymbol{\varepsilon}'_{\!0}), \cdot\rangle\boldsymbol{\Delta}_0(\boldsymbol{\varepsilon}'_{\!0})\\
    &= \sum^\infty_{i=1}\boldsymbol{\tilde{\Phi}}_i\mathscr{C}_{\boldsymbol{\varepsilon}'}\boldsymbol{\tilde{\Phi}}^\ast_i + \boldsymbol{\Delta}_0\mathscr{C}_{\boldsymbol{\varepsilon}'}\boldsymbol{\Delta}^\ast_0.
\end{align*}
As $\boldsymbol{\tilde{\Phi}}_0=\boldsymbol{\Delta}_0$, this establishes \eqref{Representation of covariance operator ARMA w.r.t. covariance operators of errors}. Finally, \eqref{covariance op of epsilons} holds because $(\varepsilon_k)$ is a $T$-PCWWN.
\end{proof}

\smallskip

\begin{proof}[\textbf{Proof of Proposition \ref{Prop: Weak dependence}}] From the linear process representation \eqref{lin process representation of multivariate efARMA(1,1)}, the definition of $L^p$-$m$-approximability, and arguments analogous to those used in the proof of Proposition~\ref{Prop: Finite higher moments}, it follows that for any $m\in\mathbb{N}$ and $\tau \geq 2$
\begin{align*}
    \big\|\boldsymbol{X}'_m - \boldsymbol{X}'^{\,(m)}_m\big\|^{\tau} &\leq \big\|\boldsymbol{\Phi}\boldsymbol{\Delta}_0 + \boldsymbol{\Delta}_1\big\|^{\tau}_{\mathcal{L}}\left(\,\sum^\infty_{j=m}\,\|\boldsymbol{\Phi}^{j-1}\|_{\mathcal{L}}\,\|\boldsymbol{\varepsilon}'^{\,(m)}_{m-j}\|\right)^{\!\tau},
\end{align*}
where 
\[
    \boldsymbol{\varepsilon}'^{\,(m)}_{m-j} \,\coloneqq\, \Big(\varepsilon^{(m)}_{m-j},~ \varepsilon^{(m)}_{m-j-1},~ \dots, ~\varepsilon^{(m)}_{m-j-(T-1)}\Big)^\top.
\]
Due to the fact that $\boldsymbol{\varepsilon}'^{\,(m)}_{m-j}\stackrel{d}{=}\boldsymbol{\varepsilon}'_{m-j}$ for all $j,m$, Assumption~\ref{Assumption: nu-th moments of errors}\,$(\tau)$ yields (see also the proof of Proposition~\ref{Prop: Finite higher moments})
\begin{align*}
    \Exp\!\big\|\boldsymbol{X}'_m - \boldsymbol{X}'^{\,(m)}_m\big\|^{\tau} &\leq \,a^\tau T^{\,\tau/2}\big\|\boldsymbol{\Phi}\boldsymbol{\Delta}_0 + \boldsymbol{\Delta}_1\big\|^{\tau}_{\mathcal{L}}\,\sup_k\Exp\!\|\varepsilon_{k}\|^{\tau}\left(\,\sum^\infty_{j=m-1}b^j\right)^{\!\tau}\\
    &\propto \rho^m .
\end{align*}
Hence, $(\boldsymbol{X}'_k)$ is $L^{\tau}$-$m$-approximable with geometrically decaying approximation errors, and consequently the same property holds for the fPARMA process $(X_k)$.
\end{proof}

\smallskip

\begin{proof}[\textbf{Proof of Theorem \ref{Theo:estimation errors fPAR}}] Under Assumptions~\ref{Theo:estimation errors fPAR} and \ref{As: Conv Ps-inv and Ps-id}, Eq.~\eqref{cons op matrix}, together with the definition of the estimators $\hat{\phi}_{k,1}$, directly yields Eq.~\eqref{rate 1st ln}.

Next, let $2 \leq \ell < T$. Basic manipulations combined with the triangle inequality, the operator-valued Hölder inequality, and \eqref{Matrix repr for estimation} imply
\begin{align*}
    \big\|\hat{\boldsymbol{\phi}}_\ell - \boldsymbol{\phi}_\ell\big\|_\mathcal{S} 
    &\leq \big\|\hat{\boldsymbol{\Phi}}_{[\ell]} - \boldsymbol{\Phi}_{[\ell]}\big\|_\mathcal{S}\Big\|\hat{\boldsymbol{\Phi}}^{\dagger}_{[\ell\ell]}\hat{\mathscr{P}}_{\ell}\Big\|_\mathcal{L} \\
    &\qquad + \big\|\boldsymbol{\Phi}_{[\ell]}\big\|_\mathcal{S}\Big\|\hat{\boldsymbol{\Phi}}^{\dagger}_{[\ell\ell]}\hat{\mathscr{P}}_{\ell} - \boldsymbol{\Phi}^{\dagger}_{[\ell\ell]}\mathscr{P}_{\ell}\Big\|_\mathcal{L} + \Big\|\Big(\boldsymbol{\Phi}^{\ddagger}_{[\ell\ell]}\mathscr{P}_{\ell} - \mathbb{I}\Big)\boldsymbol{\phi}_{\ell}\Big\|_{\mathcal{S}} .
\end{align*}
Next, by the definitions of $\hat{\boldsymbol{\Phi}}^{\dagger}_{[\ell\ell]}$ and $\hat{\mathscr{P}}_{\ell}$, and applying elementary arguments, we obtain
\begin{align*}
    \Big\|\hat{\boldsymbol{\Phi}}^{\dagger}_{[\ell\ell]}\hat{\mathscr{P}}_{\ell}\Big\|^2_\mathcal{L}
    &= \Big\|\hat{\boldsymbol{\Phi}}_{[\ell\ell]}\hat{\boldsymbol{\Phi}}_{[\ell\ell]}^{\ast}\Big(\hat{\boldsymbol{\Phi}}_{[\ell\ell]} \hat{\boldsymbol{\Phi}}_{[\ell\ell]}^{\ast} + \vartheta^{[\ell]}_{\!N} \mathbb{I} \Big)^{-2}\hat{\mathscr{P}}_{\ell}\Big\|_\mathcal{L}\\
    &= \sup_{1 \leq j \leq K^{[\ell]}_{\!N}}\frac{\hat{\varphi}_{\ell, j}}{\big(\hat{\varphi}_{\ell, j} + \vartheta^{[\ell]}_{\!N}\big)^2} \leq \Big(\hat{\varphi}_{\ell, K^{[\ell]}_{\!N}} + \vartheta^{[\ell]}_{\!N}\Big)^{-1}.
\end{align*}
Appropriate choices of $K^{[\ell]}_{\!N}\to \infty$ and $\vartheta^{[\ell]}_{\!N}\to 0$ as $N\to\infty$ ensure \citep[see Lemma~B.1 in][arxiv version v1]{KuehnertRiceAue2025}
\[
    \Big(\hat{\varphi}_{\!\ell, K^{[\ell]}_{\!N}} + \vartheta^{[\ell]}_{\!N}\Big)^{-1}
    = \mathcal{O}_{\Prob}\Big(\varphi^{-1}_{\!\ell, K^{[\ell]}_{\!N}}\Big).
\]
Consequently, for each $2 \leq \ell \leq T$, $\|\hat{\boldsymbol{\phi}}_\ell - \boldsymbol{\phi}_\ell\|_\mathcal{S}$, and hence $\|\hat{\phi}_{k,\ell} - \phi_{k,\ell}\|_\mathcal{S}$ for $1 \leq k < \ell$, are bounded above by the right-hand side of \eqref{rate others}.

Finally, for $\ell \leq k < T$, using the definition of $\hat{\phi}_{k,\ell}$ in Algorithm~\ref{alg:AR}, the structure of the matrix $\boldsymbol{\Phi}$, and the preceding bounds, we obtain
\begin{align*}
    &\big\|\hat{\phi}_{k,\ell} - \phi_{k,\ell}\big\|_\mathcal{S}\\
    &\quad \leq \big\|\hat{\boldsymbol{\Phi}}_{\ell,T+\ell-k} - \boldsymbol{\Phi}_{\ell,T+\ell-k}\big\|_\mathcal{S} + \sum_{m=1}^{k-1}\big\|\hat{\phi}_{m,\ell}\,\hat{\boldsymbol{\Phi}}_{\ell-m,T+\ell-k} - \phi_{m,\ell} \, \boldsymbol{\Phi}_{\ell-m,T+\ell-k}\big\|_\mathcal{S}\allowdisplaybreaks\\
    &\quad \leq \mathcal{O}_{\Prob}\!\left(\bigl\|\hat{\boldsymbol{\Phi}} - \boldsymbol{\Phi}\bigr\|_{\mathcal{S}}\right) + \sum_{m=1}^{k-1}\big\|\hat{\phi}_{m,\ell} - \phi_{m,\ell}\big\|_\mathcal{S}\big\|\hat{\boldsymbol{\Phi}}_{\ell-m,T+\ell-k}\big\|_\mathcal{L} \\
    &\qquad\quad + \sum_{m=1}^{k-1}\big\|\hat{\phi}_{m,\ell}\big\|_\mathcal{S}\,\big\|\hat{\boldsymbol{\Phi}}_{\ell-m,T+\ell-k} - \boldsymbol{\Phi}_{\ell-m,T+\ell-k}\big\|_\mathcal{L}\\
    &\quad= \mathcal{O}_{\Prob}\Biggl(\max\Biggl\{\varphi_{\!\ell, K^{[\ell]}_{N}}^{-1/2}\,\bigl\|\hat{\boldsymbol{\Phi}} - \boldsymbol{\Phi}\bigr\|_{\mathcal{S}} ~,~ \Bigl\|\hat{\boldsymbol{\Phi}}^{\dagger}_{[\ell\ell]} \hat{\mathscr{P}}_{\ell} - \boldsymbol{\Phi}^{\dagger}_{[\ell\ell]} \mathscr{P}_{\ell}\Bigr\|_{\mathcal{L}} ~,~ \Bigl\|\Bigl(\boldsymbol{\Phi}^{\ddagger}_{[\ell\ell]} \mathscr{P}_{\ell} - \mathbb{I}\Bigr)\boldsymbol{\phi}_{\ell}\Bigr\|_{\mathcal{S}}\Biggr\}\Biggr).
\end{align*}
All claims are thus established.
\end{proof}

\smallskip

\begin{proof}[\textbf{Proof of Corollary \ref{theo:cons fPARMA}}] The result follows immediately from Theorem~\ref{Theo:estimation errors fPAR} and the definitions of the involved operators and estimators.
\end{proof}

\section{Concluding remarks}\label{Sec: Conclusion}

Time series data often exhibit periodic behavior. In the univariate and multivariate settings, such features are commonly modeled using periodic autoregressive moving average (PARMA) models, which are well established in the literature. More recently, functional periodic autoregressive (fPAR) models with values in separable Hilbert spaces have been introduced; however, existing work has focused primarily on their structural properties. A rigorous estimation theory for fPAR models, including consistency results, as well as a formal definition and analysis of functional PARMA models, has so far been lacking.

This paper introduces functional periodic autoregressive moving average (fPARMA) models on separable Hilbert spaces. The proposed framework extends functional ARMA models by allowing the innovations' distribution, as well as the autoregressive and moving-average orders and operators, to vary periodically across seasons. The focus of this work is primarily theoretical. We rigorously define the model, establish its well-definedness, and characterize the associated covariance structure. We derive sufficient conditions for cyclostationarity and period correlatedness, as well as for the existence of finite moments and weak dependence. The weak dependence results yield a central limit theorem under suitable assumptions and provide the foundation for consistent parameter estimation. Moreover, we develop consistent Yule–Walker-type estimators for the fPAR operators. To address the inherent ill-posedness of the estimation problem, we employ Tikhonov regularization and impose Sobolev-type smoothness conditions, which allow us to derive explicit convergence rates. Finally, we discuss the estimation of fPARMA operators in a specific setting.

Beyond the scope of the present work, future research may consider extensions of the estimation methodology to more general fPARMA models and examine empirical applications of the proposed framework.

\paragraph*{Acknowledgements} Parts of the paper were written while the first author was employed at the University of California, Davis. 

\paragraph*{Funding} The first author was partially supported by TRR 391 \textit{Spatio-temporal Statistics for the Transition of Energy and Transport} (Project number 520388526) funded by the Deutsche Forschungsgemeinschaft (DFG, German Research Foundation). 

\bibliography{fPARMAref}

\end{document}